# A comprehensive review of electricity storage applications in island systems

Georgios N. Psarros[1], Pantelis A. Dratsas & Stavros A. Papathanassiou

*School of Electrical & Computer Engineering, National Technical University of Athens, Greece*

## Abstract

Electricity storage is crucial for power systems to achieve higher levels of renewable energy penetration. This is especially significant for non-interconnected island (NII) systems, which are electrically isolated and vulnerable to the fluctuations of intermittent renewable generation. The purpose of this paper is to comprehensively review existing literature on electricity storage in island systems, documenting relevant storage applications worldwide and emphasizing the role of storage in transitioning NII towards a fossil-fuel-independent electricity sector. On this topic, the literature review indicates that the implementation of storage is a prerequisite for attaining renewable penetration rates of over 50% due to the amplified requirements for system flexibility and renewable energy arbitrage. The analysis also identifies potential storage services and classifies applicable storage architectures for islands. Amongst the available storage designs, two have emerged as particularly important for further investigation; standalone, centrally managed storage stations and storage combined with renewables to form a hybrid plant that operates indivisibly in the market. For each design, the operating principles, remuneration schemes, investment feasibility, and applications discussed in the literature are presented in-depth, while possible implementation barriers are acknowledged. The literature on hybrid power plants is mainly focused on wind-powered pumped-hydro stations. However, recently, PV-powered battery-based hybrid plants have gained momentum due to the decreasing cost of Li-ion technology. On the other hand, standalone storage establishments rely heavily on battery technology and are mainly used to provide flexibility to the island grid. Nevertheless, these investments often suffer from insufficient remunerating frameworks, making it challenging for storage projects to be financially secure.

**Keywords:** *electricity storage, hybrid power stations, renewable generation, battery storage, pumped-hydro storage*

## 1. Introduction

The pathway towards the independence of non-interconnected island (NII) power systems from fossil fuel involves the massive implementation of variable renewable energy sources (RES), [1]. However, the electrical isolation, limited size, and low inertia of islands render them vulnerable to the disturbances emanating from the stochasticity of renewable generation, threatening the grid stability and the uninterrupted electrification of end users ([2,3]). As a result, the available room for NII to accommodate renewables is restricted to avoid the consequential adverse effects deriving from the variability of traditional technologies (wind farms and photovoltaics), leaving space for RES participation in the energy mix only up to 25%, [4,5].

To surpass the obstacles of RES intermittency and achieve higher renewable penetration levels, several actions have been suggested in the literature, including the deployment of storage facilities ([6–8]), the promotion of policies incentivizing demand-side management and implementation of smart end-use solutions ([9]), as well as sophisticated generation scheduling techniques that allow for the

[1] Corresponding Author

e-mail addresses: gpsarros@mail.ntua.gr (G. N. Psarros), pantelisdratsas@mail.ntua.gr (P. A. Dratsas), st@power.ece.ntua.gr (S. A. Papathanassiou)

better exploitation of existing conventional production assets ([10–12]). Amongst the available flexibility resources to support the integration of intermittent RES in NIIs, electricity storage is possibly the only feasible and realistic alternative ([13–15]), as it is characterized by ease of implementation compared to the demand-side management alternative, which is associated with the need for modifications on the load profile of end-users, [16].

There are numerous surveys in the literature focusing on different storage perspectives, including storage technical and economic characteristics [17–20], technologies available and their maturity levels for immediate or future implementation [21,22], technological and economic prospects [23,24], and system storage requirements [25]. Such research is either addressed to storage applications in large, interconnected power systems or presents findings unrelated to the size and characteristics of the grid but rather focused on the specific features of the technology.

Several review papers on island systems include storage-related aspects as a side topic. Specifically, the review of [26] recognizes the storage technologies proposed for specific isolated systems and focuses on the demand-side management alternatives that could potentially find implementation in NIIs. In [26], batteries and pumped-hydro storage have been identified as the leading storage technologies for islands, with the former effectively applicable to small and medium size system and the latter to large systems with natural reservoirs. In [16], the generation scheduling principles for islands experiencing high RES penetrations are investigated, showing that storage is an effecting substitute for thermal generation in terms of systems' dynamic behavior when targeting deep decarbonization. A salient outcome of [16] is that the implementation of a robust storage management strategy can warrant the secure operation of island systems, even in scenarios characterized by full-scale RES integration. The review of [9] highlights the significance of storage as a necessary component for the island's smartification. This analysis entails an enumeration of the spectrum of storage technologies primed for prospective integration into NII systems, along with a brief description of the pivotal services that storage technologies offer to improve the operation of island systems. Similarly, in [6], a short survey on storage technologies available in islands is also performed, concluding that combinations of storage technologies with different energy capacity characteristics can better serve the energy and balancing needs of islands. In [27], the difficulties of increasing the participation of variable renewables in the energy mix of French islands. The author of [27] claims that these obstacles can be facilitated through the concurrent integration of sophisticated RES production forecasting tools, optimal generation management techniques, and the strategic deployment of energy storage systems. The survey of [28] suggests that electricity storage stations are the best candidates for furnishing critical frequency and voltage support to the grid during high RES penetration intervals. Nevertheless, the paper acknowledges that the adoption of these solutions entails amplified capital investments; therefore, a thorough feasibility analysis is imperative to determine whether their installation aligns with the system's benefits.

Review articles dedicated to storage facilities in NIIs typically delve into the challenges of attaining high RES penetration conditions. Within this context, storage is often leveraged to increase the control over intermittent renewables and to mitigate curtailments. In [29], the hybridization of wind generation with the introduction of pumped hydro storage systems is investigated. The findings indicate that these integrated storage and RES facilities have the potential to facilitate increased renewable penetration levels in islands without compromising system stability. In the same vein, the review of [30] delves into the opportunities for introducing hybrid renewable and storage stations, investigating their techno-economic viability across systems of varying sizes in the Aegean Archipelago, Greece. The review's findings conclude that such implementations would generally prove economically advantageous for the systems under study. Batteries emerge as a more favorable choice for smaller islands, whereas pumped hydro technology prevails in the larger NII systems, particularly when substantial energy storage capacities are explored. The review presented in [31] analyzes hydrogen storage installations and fuel cell facilities to improve the integration of variable renewable generation into European island systems, encompassing countries such as Spain, Denmark, Greece, and Portugal. Analysis conclusions suggest that opportunities for hydrogen-based solutions may be viable; however, the establishment of new legislative and regulatory frameworks and technical standards is required for hydrogen to be viably incorporated in NII systems. In [32], a pre-feasibility study is performed for numerous Greek islands to assess the economic viability of hybrid wind-pumped-hydro-storage installations, using as a metric the avoided system variable costs emanating from the operation of thermal units. Results show that substituting thermal generation with wind-pumped-hydro power plants can justify hybrid investments when evaluating the savings in fuel costs. On the other hand, the review of [33] follows the timeline of developing the hybrid wind-pumped-hydro-storage project in El Hierro island, gathering the obstacles obtained from its real-world application that preventing the effective operation of the plant under very high RES penetration conditions, as initially planned. More specifically, one major problem of the El Hierro hybrid plant is detected in the alternations of the project’s technical specifications during construction, necessitating the operation of thermal generators for grid stability. As a remedial strategy, the author suggests augmenting the energy capacity of the upper reservoir and incorporating photovoltaic systems to enhance RES generation on the island.

Contrary to the existing review papers in this field, which have extensively covered storage technologies, their maturity levels, cost trends, and specific case studies, this work focuses on the practical applications of storage within island power systems. It aims to identify potential architectures for system-level storage integration, whether used standalone or combined with renewable energy sources. The paper delves into the remuneration mechanisms envisioned for such storage facilities, investigates the financial feasibility of these investments, recognizes potential technological or regulatory barriers that may hinder storage deployment, and highlights opportunities while proposing future research directions. The value of storage as an instrument to achieve high-RES penetration levels in islands is also discussed, reviewing several available articles investigating RES penetrations from approximately 10% to 100%. Additionally, the services provided by storage in NIIs systems, and the respective storage designs available are recognized. The review eventually emphasizes the two predominant storage typologies for island applications; the centralized storage concept, where storage operates independently of renewable installations, and a hybrid concept, in which storage and renewables cooperate to inject controllable RES energy into the island grid.

A detailed breakdown of the methodology adopted and statistics concerning the papers included in this review paper can be found in Appendix A. In brief, to conduct this study, a review of several scientific and institutional documents was undertaken, encompassing a wide range of material, including research and review articles, conference papers, reports, and books that focused on various aspects of island power systems. Out of this process, 195 documents met the criteria for inclusion in this review paper, with a primary focus on topics related to storage integration and alternative storage architectures in NII systems. Papers dealing with the pivotal role of storage towards high RES penetration levels in islands were also explored. Significant weight was given to research performed within the last decade, as it is more likely to present the latest models and methods relevant to the subject under evaluation.

The reminder of this paper is organized as follows: Section 2 highlights the storage value for island systems targeting high-RES penetration levels, recognizes the services storage offers to island grids, and classifies available storage designs. Sections 3 and 4 present the fundamentals of the hybrid storage-RES designs and centralized storage concept, respectively, identifying the operating principles, remuneration policies, and applications of each. Section 5 summarizes the key findings of this survey and gives future research directions. Appendix A describes the methodology adopted and furnishes statistics for the papers of this review. Appendix B provides a short description of the main findings and comments for the papers investigated for the two main storage architectures identified by the review process. Appendix C includes the abbreviations of the paper.

# 2. The value of storage to achieving high renewable penetrations in islands

## 2.1. The pathway towards decarbonization

Several available studies have explored the pathways for the NII power systems toward a lower-emission, sustainable electricity sector, seeking independence from imported fossil fuels ([34]). Studies in Table 1 target island systems pursuing high RES penetration levels. Nevertheless, from Table 1, it can be inferred that the definition of "*high-RES penetration levels*" term is debatable, varying significantly between available papers. Several factors may affect the categorization of RES penetration levels as "*high*", including (a) the maturity of the power system in terms of already integrated RES in the energy mix, (b) the year each study has been performed, (c) the availability of primary renewable resources, (d) the decarbonization goals set, and (e) the size of the island. In this vein, [35–40] characterize "*high*" RES penetrations between 10% and 50%, while, conversely, [41–55] classify as "*high*" renewable penetrations in the range of 50% to 100%.

Notably, for the large power system of Cyprus island, annual RES penetration shares in the order of 11.5% ([35]) had been considered sufficiently high to reduce gas pollutant emissions ($CO_2$, $SO_2$, $NO_x$, $N_2O$) and to comply with the European Directives of the time the study was performed. For the NII of [36], where the primary objective was to evaluate storage integration, RES penetration levels of around 16.5% were investigated. RES penetration levels between approximately 40% and 50% have been examined in [37–40]. In these papers, various islands with diverse demand profiles were examined, with the increased share of renewables in their energy mix contributing to goals such as energy independence [38,40], variable operating cost reduction [39], gas pollutants emissions reduction [37], or a combination of the above targets.

The motivation for targeting annual RES penetration levels of up to 100% is the decoupling of the electricity sector from fossil fuel imports ([44–47,53–56]) and the reduction of $CO_2$ emissions ([4,57]). Although the pathway toward such ambitious decarbonization targets has been intensively investi-gated in the literature ([41,42,48–55,58–61]), the economic feasibility of achieving annual RES penetration levels above 50% is still questionable, as arises from [3,62]. The main inhibitory factors preventing the deep decarbonization of island systems are related to the amplified investment costs of new RES and storage investments ([42,48–51,55]) in tandem with the excessive levels of installed renewable capacities required to achieve renewables shares above 70%, which can be as high as six to seven times the peak demand requirements of the NII ([42,44,48–51,55]), affecting the final cost of electricity ([4]). For instance, in [63], an examination of 147 insular power systems in the Philippines revealed that the RES-battery portfolio qualifies as cost-optimal. The analysis led to achieving an average of 58.58% RES across all 147 islands, with a maximum of 70% for certain islands. However, one of the examined cases reached renewable penetrations near 100%. In [56], several policies were designed to promote RES penetration in 31 Caribbean island states, including investment and tax incentives, feed-in tariff schemes, net-metering, and net-billing schemes. However, the analysis did not focus on deep decarbonization, given that the states investigated still face deficient renewable installations, leading to average RES penetration rates of ca. 5% for all examined islands. The transition of several Caribbean islands towards a 100% sustainability target has been investigated in [64], exploiting the ocean thermal energy conversion technology, battery storage, and desalination facilities. Results revealed that attaining a 100% renewable penetration goal in the electricity sector might be feasible for some islands, leading to lower electricity costs than those anticipated if they were to be electrified by fossil fuels, yet, once again, such an outcome could not be generalized for the entire cluster.

On the other hand, small isolated systems, with peak demand requirements in the order of a few tens of kWs, can pave the way towards 100% RES penetration levels, given that their small size favors the deployment and testing of innovative pilot projects at a relatively low cost. Such case studies can be very small islands or isolated communities, such as in [65,66], seeking to reach 100% RES penetration levels and achieving green island targets. Eventually, both hydrogen storage and batteries are deemed necessary in such applications, with the latter acting as the balancing parties for the mismatches between RES production and demand requirements in short timescales.

Table 1 also shows that attaining annual RES penetration levels beyond 25%, necessitates more than the sole integration of variable renewables, regardless of the size of the NII under investigation or the simulation tool employed–be it custom or commercially available. Specifically, to achieve penetrations up to 50%, the installation of storage (batteries, fuel cells, pumped storage - [36,37]) or the deployment of dispatchable renewables, such as biomass/biogas stations, solar thermal, small hydropower plants ([38,40]), or a combination thereof ([39]) becomes imperative. However, the accomplishment of even higher RES penetration levels (>50%) is fully integrated with the development of storage stations ([41,42,48–52,54,55]). In such cases, storages are expected to provide the required flexibility to the NII systems ([36,50,67–75]), enabling them to effectively release their RES penetration margins ([55]). This observation is verified by [76], where the Huraa Island in the Maldives is investigated, with a peak load demand no higher than 0.4 MW. Authors in [76] conclude that the system might achieve RES penetration levels up to 53% without storage. Nevertheless, surpassing 53% and reaching levels up to 96% requires the introduction of battery energy storage stations to handle the variability of the anticipated considerable renewable installed capacity, estimated as high as seven times the load peak.

Table 1: NII power systems characteristics and achievable annual RES penetration levels.

| **Ref.** | *Island system* | *Country* | *peak demand* [MW] | *RES technologies* | *Storage technologies* | *Achievable RES penetration* | *Simulation tool* |
|---|---|---|---|---|---|---|---|
| [35] | Cyprus | Cyprus | 930 | PVs, wind, solar thermal, biomass | no | **11.5%** | undefined |
| [36] | unnamed *(Azores Archipelago)* | Portugal | 35 | wind | batteries | **up to 16.5%** | custom *(opitmization)* |
| [37] | Pulau Ubin | Singapore | <1 | PVs, wind, solar thermal, biomass | fuel cell | **20%-40%** | TRNSYS 17 |
| [38] | Pantelleria | Italy | 10 | PVs, wind, solar thermal, geothermal, waste | no | **10%-50%** | NEPLAN |
| [39] | Crete | Greece | 600 | PVs, wind | pumped storage | **up to 37%** | custom *(opitmization)* |
| [62] | Baltra and Santa Cruz | Ecuador | 7.26 | PVs, wind | batteries | **up to 39%** | HOMER pro |
| [40] | Pantelleria | Italy | 10 | PVs, wind, geothermal | no | **up to 46%** | undefined |
| [41] | Ikaria | Greece | 9 | PVs, wind | pumped storage | **>50%** | custom |
| [63] | 147 Philippine off-grid islands | Philippines | <1 (for 122 islands) >10 (for 5 islands) | PVs, wind | batteries | **up to 70%** | HOMER pro |
| [42,48] | El Hierro | Spain | 6.5 | wind | pumped storage | **up to 70%** | custom |
| [43] | Faial | Portugal | undefined | PVs, wind, geothermal | batteries | **up to 75%** | custom *(opitmization)* |
| [61] | Astypalaia | Greece | <2.5 | wind | batteries | **up to 82%** | custom |
| [49] | 5 Aegean islands *(Rhodes, Lesvos, Chios, Karpathos, Patmos)* | Greece | 5 - 200 | PVs, wind | batteries | **up to 85%** | custom *(opitmization)* |
| [46] | El Hierro | Spain | 7.5 | PVs, wind | pumped storage, batteries | **up to 85%** | TRNSYS |
| [50] | Aghios Efstratios | Greece | 0.3 | PVs, wind | batteries | **>85%** | custom *(opitmization)* |
| [51] | 6 islands *(Streymoy, Aruba, Sumba, Rhodes, Gran Canaria, Rarotonga)* | Faroe Islands, Aruba, Indonesia, Greece, Spain, Cook Islands | 25, 122, 7, 213, 548, 4 | PVs, wind | pumped storage, batteries | **up to 90%** | custom (*Matlab*) |
| [4] | 2 islands (*Tenerife, Gran Canaria)* | Spain | 588, 506 | PVs, wind | batteries | **up to 100%** | HOMER |
| [52] | Aghios Efstratios | Greece | 0.3 | PVs, wind | batteries | **up to 100%** | HOMER |
| [44] | Faroe Islands | Faroe Islands | <150* | PVs, wind, tidal, biofuels | pumped storage, batteries | **up to 100%** | Balmorel |
| [47] | Gran Canaria | Spain | ~454 | PVs, wind, biomass, waste | pumped storage | **up to 100%** | EnergyPLAN |
| [64] | Caribbean islands | Caribbean nations *(several)* | 2 up to 3685 | PVs, wind, geothermal, ocean thermal | batteries | **up to 100%** | custom (*Python*) |
| [53] | Reunion | France | >500 | PVs, wind, biomass, hydro, wave | no | **100%** | TIMES |
| [54] | Reunion | France | >500 | PVs, wind, biomass, hydro, wave | batteries | **100%** | TIMES |
| [55] | Porto Santo | Portugal | 5.6 | PVs, wind | fuel cell *(hydrogen)* | **100%** | custom *(H2RES)* |
| [57] | Baltra & Santa Cruz | Ecuador | 13.4 | PVs, wind | pumped storage, batteries | **up to 100%** | HOMER Pro |
| [45] | Ometepe | Nicaragua, | 4 | PVs, wind | pumped storage, batteries | **100%** | custom *(opitmization)* |
| [65] | unnamed | Italy | <1** | PVs, wind, small hydro | batteries<br>fuel cell *(hydrogen)* | **100%** | custom *(opitmization)* |
| [66] | Juara village Tioman Island | Malaysia | <0.12 | PVs, wind | batteries<br>fuel cell *(hydrogen)* | **100%** | HOMER |
| [59] | Wang-An | Taiwan | 1.3 | PVs, wind, biomass, wave | yes *(not specified)* | **100%** | EnergyPLAN |
| [60] | 2 islands *(King Island, Molokai)* | Tasmania, Hawaii | 2.5, 5 | PVs, wind | batteries<br>flywheels | **>60%, up to 100%** | HOMER pro |

* *Estimate combining data from Figure 4 of the paper and installed capacities of system assets.*

** *Concluded from the analysis. Not directly specified within the paper.*

## 2.2. Services provided by storage in islands

Undoubtedly, energy storage stations (ESS) are vital for the electricity sector of NII to move to penetrations of renewables over 50%. As can be inferred from Table 1, pumped hydro storage (PHS) and battery energy storage (BES) technologies dominate the landscape of actual grid-scale applications for island systems. Pumped hydro was the default technology of choice up to some years ago due to its technical maturity and the hydro resources available in certain islands, [41,77]. However, batteries have already emerged as a realistically feasible storage technology for their ease of deployment, declining cost, and ubiquity ([36,50,78–83]). Other storage technologies, including compressed air energy storage, thermal energy storage, hydrogen storage systems, flywheels, and supercapacitors, may be the subject of attention in the literature. However, as will arise later in this paper and as indicated by Tables B1 and B2 in Appendix B, such energy storage solutions have been examined to a limited extent for island applications.

According to the relevant literature ([84–88]), electricity storage stations can be classified into several categories based on different criteria, such as their size, maturity of storage technology, available services they may provide to the power system, etc. A fundamental classification of ESS distinguishes mid- to long-duration storages, capable of supporting bulk energy transactions in daily or seasonal arbitrage, from short-duration storage stations[2]. The latter can serve as system-level com-ponents contributing to system balancing and other technical services (e.g. voltage/frequency support, inertia emulations, short circuit current), or they might be distributed in the MV-LV grid and behind-the-meter in consumer facilities.

[2] In this paper, long-duration storages are defined as those with the capacity to store substantial amounts of energy and engage in multi-daily or seasonal arbitrage, utilizing energy-to-power ratios exceeding 10-h ([194,195]). Short-duration storage stations are primarily employed for power-intensive applications and potentially limited daily arbitrage, characterized by energy-to-power ratios up to 4 hours. This upper limit is notably defined by battery technology. Storages featuring ener-

The benefits from both long- and short-duration ESS can prove to be instrumental for isolated power systems characterized by low inertia and lack of interconnections to larger systems, which, in turn, create amplified needs for regulation services, particularly when exposed to high penetrations of intermittent renewable production. Integrating energy storage can alleviate several operating and management difficulties that NIIs face during real-time operation. Some of the tangible benefits from the deployment of storage, common to a large extent for all power systems, isolated and interconnected, are the following:

- **Energy-oriented services** predominantly provided by long-duration storages. These include energy arbitrage, [67,74,75], (energy transfer between the low-demand, low-price hours in the valley of the load curve and the high-demand, high-price hours during evening peaks), as well as support to the increased participation of intermittent renewables, [36,50,68,69], (storage absorbs any surplus of renewable production which would be curtailed otherwise).
- **Fast response and active power reserves** ([70–73]). To ensure the secure operation of any power system, it is essential to maintain operating reserves to deal with disturbances ([89]), such as RES fluctuations or loss of online generating units. Fast response ESS (e.g. batteries, flywheels, etc.) incorporate the technical capability to instantaneously respond to such events, literally within a fraction of a second and much faster than any conventional thermal unit. Such services are most critical in small isolated systems, exposed to increased RES volatility levels and large-scale disturbances, compared to the size of the system. Among the various frequency-related services, fast reserves are by far the most critical for the integrity of the systems and the possibility of incorporating large RES capacities ([12,90–93]).
- **Contribution to voltage regulation** ([94,95])**.** Voltage regulation and management of reactive power balance is another area where ESS may contribute substantially when suitably located in the island grid. Such a service becomes critical under high-RES penetration conditions, where conventional generators, usually in charge of voltage control, are massively substituted by renewables with limited voltage regulation capabilities. In this case, storage can effectively provide these services, as inferred by the analysis in several studies available ([28,96,97]).
- **Resource adequacy-oriented services** ([98–101])**.** Either operating alone or combined with intermittent RES technologies, storage stations might provide firm capacity to the system subject to their power and energy capacity limitations, substituting thermal generators traditionally assigned with this task. Such security of supply services provided by storage are essential for power systems transitioning to low-carbon emissions, whose energy mix gradually displaces fossil fuel-fired assets in favor of variable renewable generation and storage.
- **Black start services** ([102–105])**.** Storage stations with grid-forming capabilities and sufficient energy reserves may actively contribute to system restoration after black-outs. This functionality is significant in small NII systems of low inertia, where black-out events occur more frequently than in robust interconnected grids.

gy-to-power ratios above 4-h and below 10-h belong to the mid-duration category, including small pumped-hydro stations or large batteries (6-h).

## 2.3. Recognizing electricity storage typologies in islands

So far, the literature review has evident the role of storage facilities in providing pivotal services for the transition of island systems to higher penetration of renewables. It is worth noting that storage stations can be deployed in island power systems under various management and operating designs. The suitability of each storage design for implementation relies upon multiple factors such as market structure, island size, the provisions of the prevailing regulatory framework, etc. It is also essential to recognize that adapting to different designs often introduces management rules that might limit the ability of storage units to provide their full spectrum of services to the grid.

The main ESS deployment alternatives met in the literature for island systems are illustrated in Figure 1. Regardless of their technology, storage stations can be implemented in islands as (a) standalone, front-of-the-meter plants, (b) combined with traditional renewables within a virtual power plant (VPP), (c) embedded in behind-the-meter applications coupled with RES, or (d) implemented within the facilities of end consumers/prosumers.

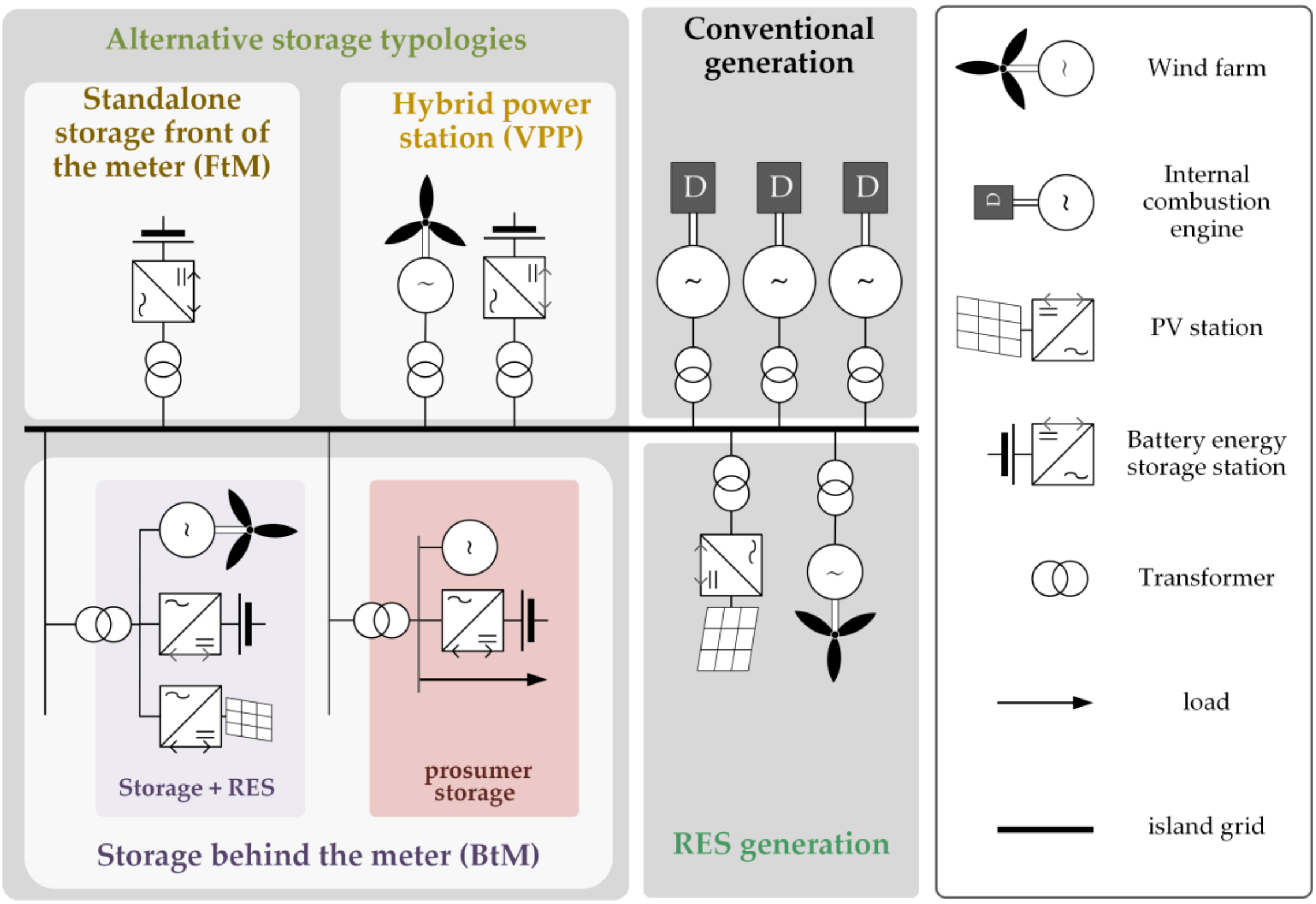

*Figure 1: Alternative deployment paradigms of storage in NII power systems.*

Standalone storage applications are usually directly supervised and centrally managed by the island System Operator (SO); thus, they are suitable for providing all kinds of system services described in Section 2.2, or local network support, as needed and requested by the SO. The application of such storage architecture is quite appealing for power systems lacking organized electricity markets, which is the case in most medium and small islands, where a single entity, usually the SO, is responsible for dispatching and controlling all system assets.

Combinations of storage and renewables to form a VPP have also been examined widely in the is-lands' literature under the hybrid power station (HPS) concept. HPS plants are governed by specific rules that allow for the coordinated participation of the storage and RES constituents of the station in the electricity market of the island as a single entity. HPS are generally self-scheduled plants responsible for internally dispatching their generation and storage components. In this sense, the components of the HPS are not directly managed by the SO, as is the case with the standalone storages, but by a third party, usually the station's owner. HPS are designed to primarily provide energy-oriented services to the island grid, targeting to displace conventional generation during the peak-demand time intervals of the day in favor of renewable production originating from HPS facilities.

Behind the meter applications, e.g., in consumers, present largely similar possibilities in island and in mainland system context, being principally driven by the use case of the end-user. In fact, island applications may lack in potential compared to similar cases in large systems, owing to the fact that organized electricity markets offer opportunities for additional revenues, which may not be the case in the islands. This is the reason why the literature of behind the meter storage applications in consumers specifically for islands is unsubstantial.

In the same vein, storage embedded behind-the-meter of RES stations is also of limited interest for island systems. Only a few studies available ([8,106–108]) investigate the potential of using storage (batteries) behind-the-meter of wind farms (WFs) in order to better exploit the wind regime which otherwise would be curtailed during specific time intervals of RES congestion in the grid. The main finding of these studies is that battery energy storage stations (BESS) of increased energy capacities can indeed improve the performance of the RES plant in terms of wind curtailment reduction. Nonetheless, the capital costs for implementing large storage capacities were proven exceptionally high, making it difficult for independent power producers to rationalize such investments.

Given the limited interest in the literature for behind-the-meter storage applications in island systems, this paper will not further explore such concepts. The interested reader is kindly referred to [109–111] for further information on such designs, which, however, find wider implementation in large continental systems with well-established electricity markets.

# 3. Storage coupled with RES within a Hybrid Power Station

## 3.1. Definition & Structure

The HPS design in NIIs foresees the establishment of an aggregate station comprising only storage and variable renewable generation facilities, akin to the concept of VPPs ([112–118]). Contrary to the classic VPP design, HPSs exclude conventional generation assets from their components, qualifying as purely renewable plants with enhanced dispatchability capabilities originating from the presence of storage and the rules governing their market participation. Notably, the underlying idea of the HPS design is that through effective coordination of the renewable and storage facilities, the stochasticity of HPS intermittent renewable sources will be successfully mitigated, enhancing the controllability of the entire plant, and eventually delivering dispatchable energy to the island grid. The HPS concept targets "*energy intensity*" storage installations, as it is addressed to storage stations incorporating large energy capacities, usually with energy-to-power ratios in the order of 8 hours or above[3]. HPS dispatchability attributes, in tandem with the increased energy capacities accompanying its storage assets, allow for the station to contribute to NII resource adequacy. This feature of HPS allows for deferring or even eliminating investments in thermal generation, which would otherwise be required to serve resource adequacy purposes. In this sense, HPSs aim to increase renewable penetration rates in already RES-saturated island systems that cannot accommodate further installed capacities of non-dispatchable renewable stations while concurrently providing amplified security of supply levels.

The HPS design for island systems was initially introduced by the work of [42],[48], where the El Hierro island was used as the case study to achieve a promising >70% penetration of renewables, combining wind farms and pumped hydro-storage facilities. The HPS concept was better shaped a couple of years later by the work of [119], while the main structure of the operating policies that could be applied to HPS and the management of their assets was put in order by the work of [120]. Up to 2019, the leading storage technology investigated in the literature for HPS applications was the pumped-hydro stations ([32,119–123]), which were perceived to be the most suitable and probably the sole economically feasible storage alternative. Thus, pumped hydro technology, traditionally associated with increased energy capacities up to 100-h, was ideal for the storage-RES aggregation to be classified as HPS, aligning with the amplified contribution to adequacy prerequisite. Therefore, the interest of the literature for HPS was solely focused on pumped-hydro stations combined with wind farms, examining thoroughly several aspects of this architecture, including the value they add to the island system ([32,41,123]) from a holistic viewpoint (RES penetration, system economics), the dynamic behavior of the system and the respective fast response capabilities of HPS configurations ([77,124,125]), the sizing of HPS individual components ([121,126]), as well as issues related to the participation of the aggregate plant in the electricity market of the island ([41,122]) according to the rules and constraints governing its operation.

The dominance of the pumped-hydro storage in the HPS concept began to wane during the last few years when literature started to investigate batteries as an alternative storage technology for incorporation in the hybrid plant [61,82,99]. Such a transition is mainly associated with the current investment interest in batteries, originating from the decreasing capital expenditure trend for utility-scale BESS of Li-ion technology ([127–130]), the reliability of battery energy storage equipment that has improved the maturity level of the technology, the amplified battery lifespan over the last years, the high roundtrip efficiency and the ease and speed of deployment of the relevant facilities. Another aspect contributing to the shift towards batteries is the lack of suitable island geomorphological locations to develop pumped-hydro stations. The gradual transition to BESS technology might compromise storage durations to ratios below 8-h, as is the case of the 7-h HPS plant in Tilos ([131]). Such reductions in duration are associated with the relatively high energy capacity costs of batteries, which restrain the sizing of the project in favor of its financial viability.

The attention in BESS for HPS came with increasing interest for solar photovoltaics (PVs) as the primary resource of HPS's energy instead of WFs, which was the most appealing RES technology for HPS applications up to 2015 (Table B1 of Appendix B). PVs became an attractive alternative in terms of investment costs, required development times, and the possibility of implementation in smaller-scale HPSs, sectors in which WFs are at a disadvantage.

[3] The storage duration for the HPS is usually determined based on the firm capacity of the plant, which is also known as the "*guaranteed*" power. This capacity may be lower than the installed power capacity of the storage station in the portfolio, [82].

## 3.2. Operating principles & participation in islands electricity markets

The operating principles for HPSs in NII power systems are addressed in [120], [132]. For the hybrid plant to participate in the day-ahead market, its operator (HPS-O) initially submits energy offers covering the entire 24-h dispatch horizon of the day, aiming at maximizing the plant's revenue ([41,99,121,133,134]). These energy offers denote the energy availability of the plant within the next day, as estimated by the HPS-O, and can be fragmented within the hours of the day by the SO when solving the UC-ED process in the most convenient manner to serve the system cost minimization objective. HPS energy offers in the market are non-priced ([122]) and, as they are of renewable origins, enjoy dispatch priority against thermal generation ([82,119]). Further, the energy dispatched, and reserves allocated to the HPS per dispatch interval should not exceed the rated power of the aggregate plant.

HPS energy offers are generally dispatched during the peak demand hours of the load curve ([120]), aiming at minimizing daily system costs by substituting expensive peaking conventional generators (peak shaving), as qualitatively depicted in Figure 2(a). Further, the HPS are obliged to provide the so-called "*guaranteed*" energy & power, thus contributing to the resource adequacy of the NII power system. To do so, if their energy offer does not cover a "*guaranteed*" energy threshold, explicitly defined in each HPS license agreement, HPSs are eligible to absorb energy from the grid when instructed by the SO to fulfill their capacity adequacy obligations ([41,82]). Generally, this operation takes place during the low-demand hours at the valleys of the load curve (as in Figure 2(b)) on the day that resource inadequacies are foreseen.

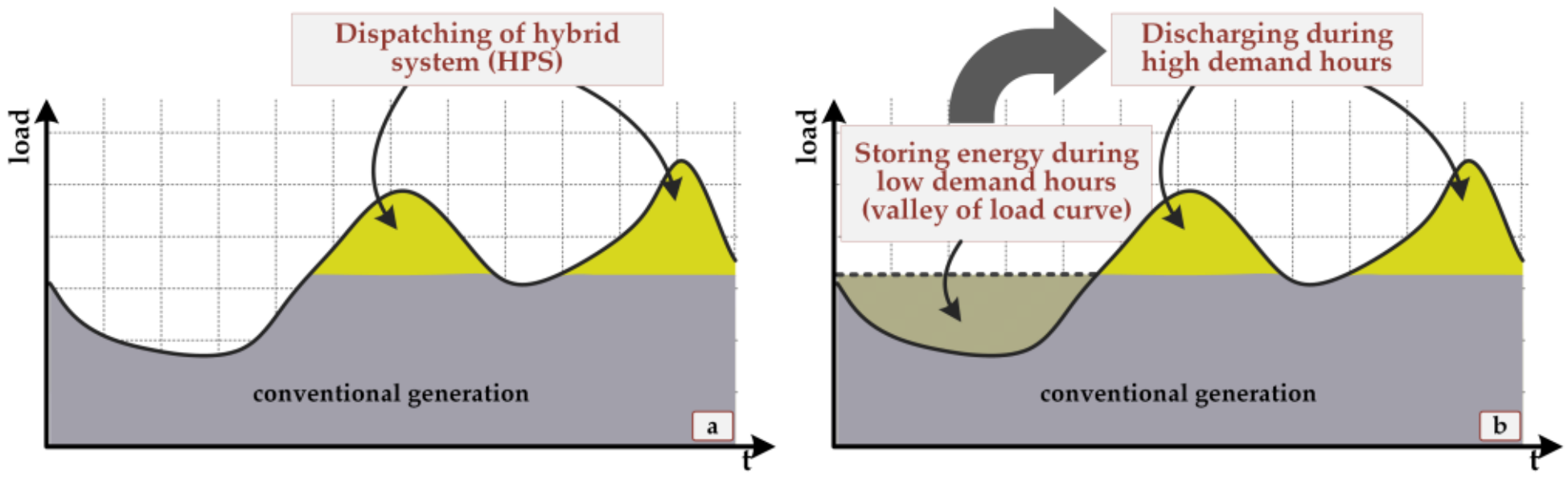

*Figure 2: (a) Dispatching of HPS energy offer within the load curve and, (b) HPS grid energy absorption and reinjection into the load curve to provide "guaranteed" energy & power.*

After dispatching the energy offers within the daily load curve, the hourly dispatch plan of the HPS is fully formed and communicated to the HPS-O. From that point, the HPS-O is obligated to comply with both energy and reserves dispatch orders ([82]), by optimally self-scheduling the internal components of the plant (renewables and storage). During real-time operation, four different operating modes are proposed in the literature and presented in Figure 3 for the self-scheduling of HPS components, principally targeting the optimal exploitation of the primary renewable HPS regime ([41,82,99,120,122,135]):

- **Mode I:** When HPS operates in this mode, the renewable generation of RES installations embedded in the HPS is stored in the plant's storage facilities. Mode I is typically enabled when RES generation exceeds the dispatch order for the examined time interval and the storage facilities have enough room to accommodate it.

- **Mode II:** In this mode, HPS renewable generation is directly injected into the island grid to meet the dispatch order, fully or partially. The level of participation of HPS renewable generation in the fulfillment of the dispatch orders might be subject to further limitations, which are related to the required reserves the HPS plant should maintain to ensure that possible fluctuations of its variable renewable resources will not affect the output of the entire plant and the quality of the power delivered to the grid.
- **Mode III:** In this mode, the renewable energy that has been previously stored in HPS storage facilities is directly injected into the grid to fulfill a dispatch order.
- **Mode IV:** In this mode, the HPS absorbs energy from the island grid after SO's command to ensure that its energy reserves are enough (up to the levels of guaranteed energy) to cover potential resource adequacy issues of the NII.

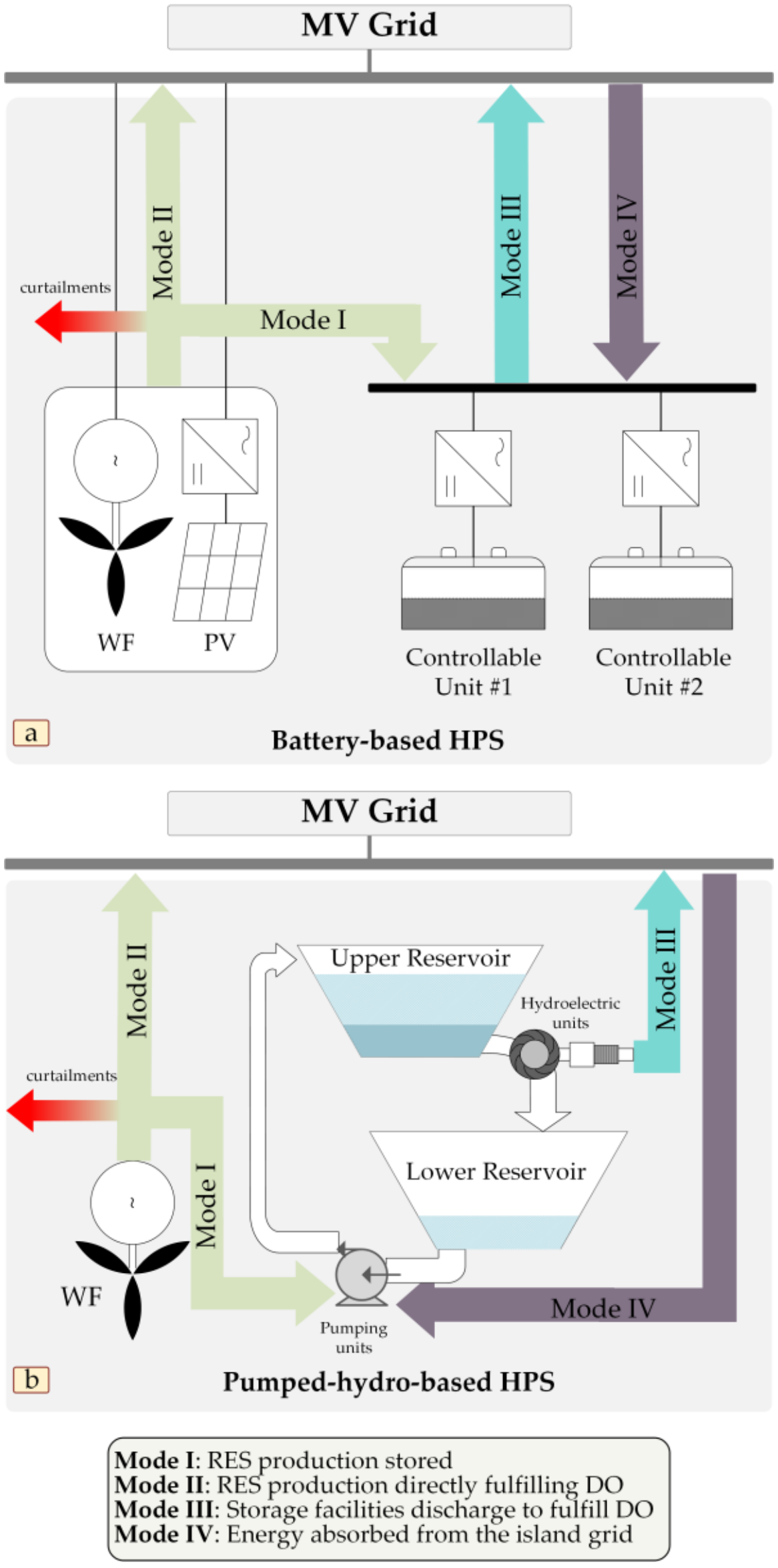

*Figure 3: Schematic of HPS and component operating modes for (a) battery and (b) pumped-hydro storage HPS technology.*

## 3.3. Remuneration policies

### 3.3.1. Initial approaches

The HPS design is a quite popular storage-RES combination alternative in the Greek islands with several applications, as Greece has a well-established legislative framework to manage this concept effectively ([136], [137]). Until recently, the remuneration fundamentals for HPS production in the Greek islands were constructed upon the avoided cost principle. According to this approach, HPSs should be reimbursed for the energy they inject into the grid and their contribution to NII resource adequacy, with the total annual revenue of each station to reflect the avoided variable and fixed cost of the respective thermal peaking units that would otherwise provide these services to the system.

More specifically, the tariffs for the hybrid energy injected into the grid for operation under mode III were supposed to reflect the average annual variable operating cost of peaking units for the island system under examination. Similarly, the tariffs for energy absorbed by the island grid to fill storage facilities when HPS operates under Mode IV were designed to capture the average annual variable operating cost of base-load units for the island system under examination. Regarding the price of renewable energy injected directly into the grid as a substitute for hybrid energy to cover dispatch orders without being circulated in storage facilities (operation in Mode II), it was calculated as the average of the hybrid energy tariff for operation in Mode III and the Feed-in-Tariff (FiT) of the respective RES technology ([122]). Apparently, the calculation of energy quantities that were to be remunerated based on the abovementioned tariffs was performed *ex-post* by the SO during the market clearing process. Finally, the remuneration framework for HPS provided for the compensation of the plants for the provision of firm capacity. Each HPS was eligible to provide firm capacity equal to HPS rated power. The remuneration for this service was designed to reflect the avoided cost of implementing a new thermal power station on the island of the same capacity as the capacity of the respective HPS.

### 3.3.2. Current remuneration scheme fundamentals

As explained earlier, the rationale behind the tariff calculation reflects the specific cost conditions prevailing in each island system. This results in different remuneration levels for HPS between islands, depending on the existing thermal fleet of each. Such tariff design does not account for plant viability or incorporate measures to prevent investment over-compensation phenomena, such as the establishment of an effective claw-back mechanism, leaving the HPS plant with either the uncertainty of investment feasibility or the potential of increased revenues.

Such distortions in the HPS remuneration have been addressed through the EU decision of [138], which has radically transformed the remuneration policies for the HPSs. The main guideline outlined in this decision is to compensate the plants in a manner reflecting their levelized cost of electricity (LCOE) rather than the avoided cost of thermal production. This approach is not tailored to the specificities of each island system but is closely associated with the RES and storage technologies incorporated into the HPS plant. To select the HPS projects qualifying as mature for deployment in island systems, the revised framework defines eligibility criteria, including the minimum energy storage duration of HPS facilities, without altering the principles governing the operation and market participation of the plants in the islands' electricity markets, as previously described.

## 3.4. Summary of HPS studies and applications

Table B1 in Appendix B provides an overview of the key literature related to the HPS concept in island systems. Figure 4 (a) and (b) display the islands under consideration for the hybrid power plant design and their respective countries. Notably, the most extensively studied NII power system is that of El Hierro in Spain, followed closely by the islands of Crete and Lesvos in Greece. The extensive research on El Hierro can be attributed to two main reasons. First, the HPS scheme was originally conceptualized for this study case. Secondly, the wind-pumped-hydro storage plant has already been implemented and is operational on the island, providing a solid foundation for further exploration.

On the other hand, most of the research on the HPS concept pertains to the Greek islands (28 out of 43 study cases refer to NIIs of Greece), where this scheme is well-shaped and recognized as an instrument to increase RES penetration levels. At the same time, the prevailing regulatory framework for Greek islands remains open to independent power producers interested in implementing an HPS plant and participating in the island markets, adhering to the rules outlined previously in this Section. These conditions establish a secure investment environment for the HPS concept, with two applications already in progress in Ikaria and Tilos islands [131,139].

The characteristics of the investigated HPS plants are illustrated in Figure 5. Notably, only four storage technologies have been explored in the relevant literature. Pumped hydro storage is the most extensively examined technology, representing 66% of the cases, followed by batteries (21%) and hydrogen storage (9%). CAES technology is only present in 4% of the cases, appearing in two papers ([140,141]). When it comes to renewable technologies, wind farms, either alone or in combination with other RES units, dominate and account for 97% of the cases studied.

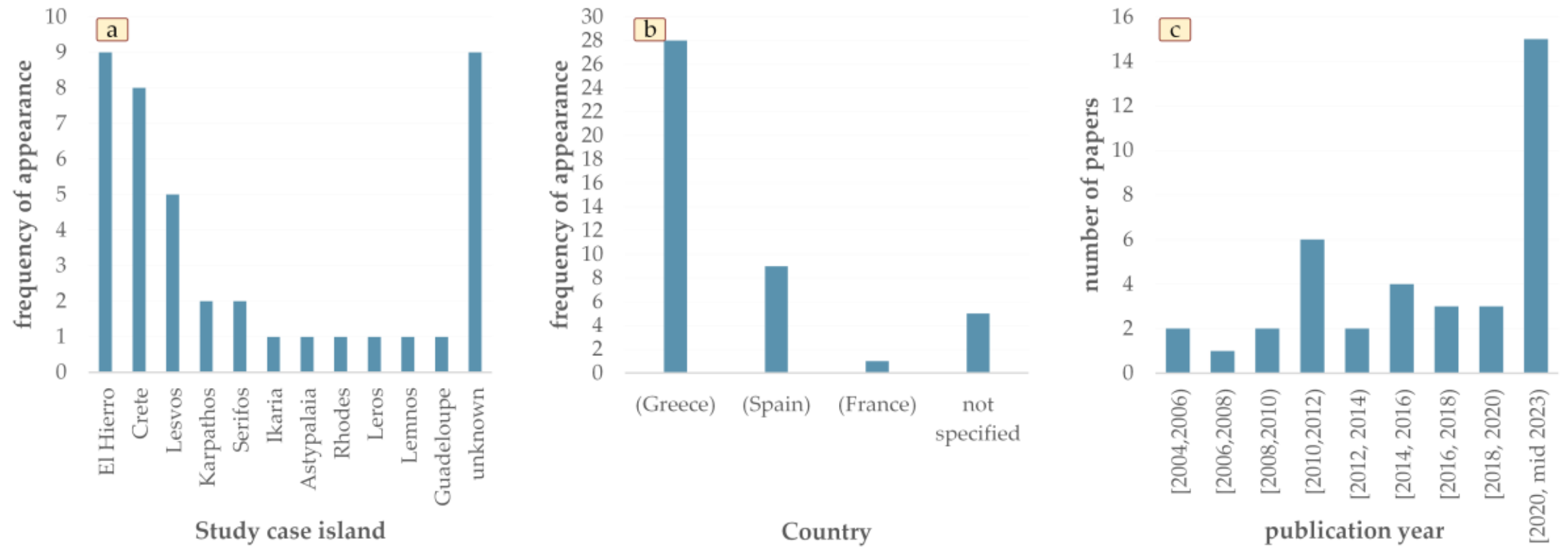

*Figure 4: (a) Study case islands (b) country they belong to for the HPS concept papers reviewed in this study and (c) publication year of the papers.*

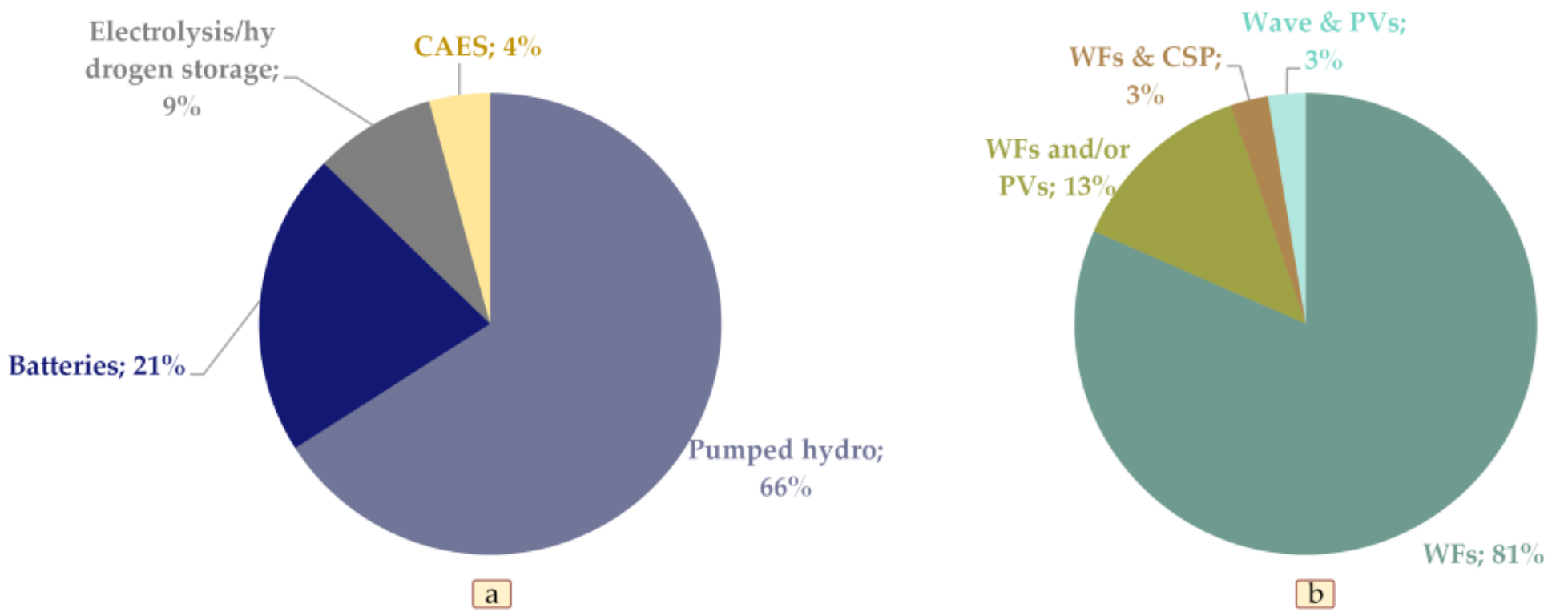

*Figure 5: (a) Storage and (b) renewable technology in the HPS papers reviewed in this study.*

The main topics addressed for HPSs are presented in Table 2. Figure 4 (c) illustrates the publication periods of the papers listed in Table 2. Given the limited literature on HPS design, most available papers have been included in this review to thoroughly capture the evolution of the concept. Nonetheless, Figure 4 (c) demonstrates that the majority of HPS papers were published within the last decade, indicating that the concept reached higher maturity levels during this period.

Evidently, the optimal sizing of the HPS components is the topic most investigated, followed by the feasibility of the relevant investments. This observation is quite important, as it reveals that while the HPS concept *per se* is well defined, the dimensioning of its constituents and the corresponding feasibility of the investment remain open issues. These factors should be *ad-hoc* accounted for in each case study island that has its own unique electrical characteristics, operating constraints, and needs. Thus, there is no one-way answer regarding the sizing and economic viability of an HPS plant, and each case should be examined independently. Should the HPS concept remain relevant in the forthcoming years, the literature on the sizing and feasibility of these stations is expected to increase further. Generation scheduling principles for NIIs in the presence of HPS plants, system stability issues, RES integration concerns, and HPS operating policies have also been investigated in several papers. Given that these research areas are highly correlated, each paper may address more than one topic at a time, as is apparent from Table 2.

On the other hand, less research has been conducted on optimal sitting, interactions with other sectors (heating), and real-time self-scheduling of HPS plants. These research areas are generally less explored, either due to their inherent complexity ([135]) or their direct focus on concerns specific to particular study-case islands, [46], [123], which cannot be clearly generalized.

*Table 2: Aim & Scope per publication dealing with HPS.*

| ***research topic*** | **Ref.** | **No.** |
|---|---|---|
| System dynamics & Frequency stability | [77][125][142][143][144][145][140][146] | 8 |
| Capacity adequacy & Capacity credit | [98][99] | 2 |
| Review | [147][141] | 2 |
| Generation scheduling | [122][148][149][82][150][119][140][146][151][152] | 10 |
| RES integration | [120][41][46][99][140][146][152][153] | 8 |
| Optimal HPS sizing | [32][42][48][61][99][119][121][126][132][152][154][155][156] | 13 |
| Investment feasibility | [32][61][82][99][123][126][132][152][155][156] | 10 |
| Sector coupling | [46][156][157] | 3 |
| Operating policies | [41][42][48][120][146] | 5 |
| Optimal HPS sitting | [123][152] | 2 |
| Market bidding strategy | [133][134] | 2 |
| Real time management | [135] | 1 |
| **Total** ***(single papers)*** | | **39** |

# 4. Centrally managed standalone storage installations

## 4.1. Definition & Operating principles

The concept of centrally managed storage facilities involves installing standalone storage units directly connected to the grid. Their operation is not associated with the generation of specific RES stations. These establishments are at the disposal of the SO to operate effectively for the benefit of the entire system (welfare maximization objective) rather than as independent assets managed by third-party entities to maximize profit. Notably, this concerns the management of storage, while ownership and technical operation responsibilities may belong either to the SO or to any other third-party remunerated by the SO for the services provided.

In this context, storages do not participate in the island's market as generation or load entities. Instead, they are treated as system-level flexibility assets designed to optimize system operation, ensure fulfillment of security constraints, maximize RES absorption, and minimize generation costs during the UC-ED processes. More specifically, a flexible centrally managed storage station, such as batteries, is anticipated to substantially enhance system operation by providing flexibility.

A fundamental advantage of this storage design compared to others is **the full controllability of the station by the SO, enabling the provision of power reserve products** as necessary to benefit the power system in its entirety without relying upon the market participation obligations of the HPS paradigm. Flexible storage assets are eligible to furnish power reserves, which are scarce in most NII power systems of small and medium size ([80,81]), provided that a sufficient energy reserve is maintained to support the corresponding reserves deployment time requirements ([158]), partially disengaging the conventional units on undertaking these tasks. This feature has proven instrumental in mitigating renewable curtailments and releasing the RES hosting capacity of the island. Specifically, it allows for the relaxation of hard technical limitations related to the operating attributes of thermal generators, i.e., must-run or minimum loading operation ([78]). Figure 6 qualitatively illustrates the reduction of RES curtailment and the corresponding increase in RES penetration during an indicative 24-h load demand profile stemming from the decommitment of conventional production for reserve provision. Note that centrally managed storages in a NII system, contrary to the HPS or behind-the-meter storage concepts, benefit all RES capacity operating on the island rather than individual facilities contributing the most to the system variable cost minimization objective ([80,99]).

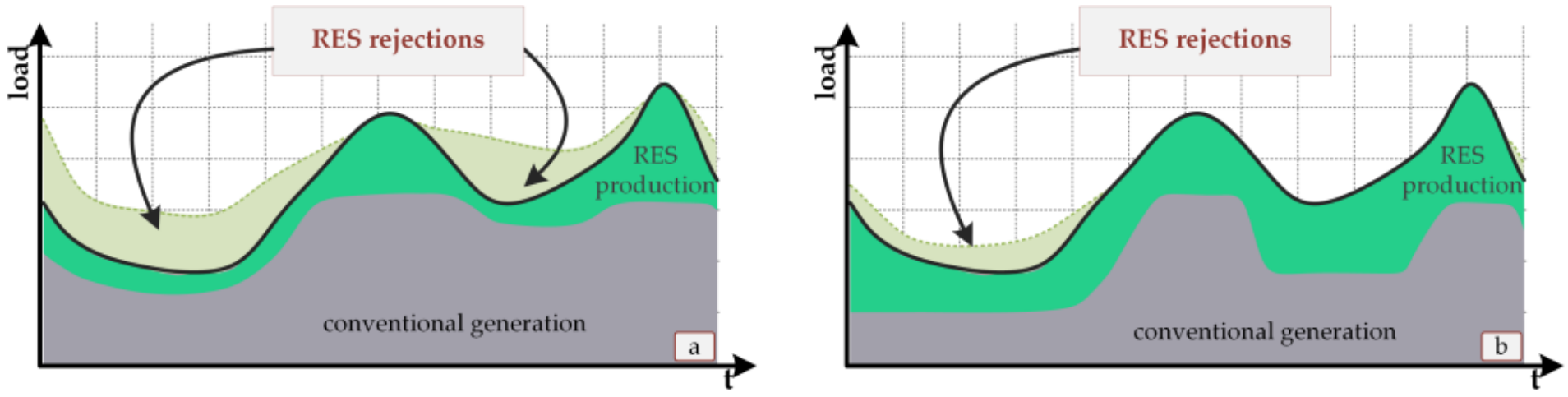

*Figure 6: RES curtailments and system operation (a) without and (b) with centrally managed storage integration.*

Centrally managed storage can also engage in energy arbitrage along with other services, [78,88]. This functionality is advantageous when the spread of system variable generation cost within the day is high enough to compensate for roundtrip losses of the storage facilities. At the same time, it requires a large enough energy capacity to transfer significant amounts of energy from the valley to the peak of the load curve, [159]. Substantial benefits are expected for NII systems incorporating thermal units with significant variable generation cost spread (e.g. low-cost base load units using heavy fuel oil and expensive peaking units burning light fuel oil). Additionally, the gradual integration of solar generation into power systems, which has gained increasing significance in recent years, emphasizes the value of the energy arbitrage functionality. This is because the highly concentrated mid-day PV generation necessitates a substantial energy shift to meet the demand peak in the evening [78,160].

Complementary to energy arbitrage, centrally managed storages can also contribute to the resource adequacy of island systems. The contribution of storage to capacity adequacy constitutes a challenging topic, with limited studies available in the literature [161–163], and even fewer focusing on the specificities of island systems [100], [101]. In [101], the firm capacity potential of BESS with different power and energy capacities is investigated under three different operating policies: cost-driven, capacity-driven and a hybrid approach that combines the principles from the former two. The study employs the state-of-the-art stochastic Monte-Carlo technique, [164], and the well-established reliability metrics of loss of load expectation (LOLE) and expected energy not served (EENS). On the contrary, in [100], the authors use a simplified but intuitive load-leveling technique to approximate the contribution of storage to capacity adequacy. Each storage configuration is attributed a firm capacity equal to the reduction of annual peak demand achieved due to the presence of storage in the island system. All studies concluded that batteries of low energy-to-power ratios, below 2-h, struggle to achieve capacity credits over 50%, regardless of the characteristics and the RES penetration levels of island systems under examination. However, batteries with higher energy capacities, over 4-h, contribute to adequacy with almost 100% of their rated power, significantly improving the system's secure operation.

Further benefits of system-level storage, which are difficult to explicitly quantify in a meaningful and generalizable manner, include the improvement of dynamic behavior and stability of the island system [71,165–167], benefits in reactive power management and voltage regulation, [94,95], possible black-start functionality [105,166], and other technical services.

## 4.2. On investments feasibility and potential remuneration schemes

While in the previously described HPS concept, investment decisions are made by the HPS producer based on a clear business case– given that the remuneration scheme is well-defined (see Section 3.3)– the feasibility of centralized storage investments in island systems remains an open issue. The relevant literature is still vague regarding the remuneration of the services provided by such stations. This is because organized markets, as provided for in continental power systems, do not thrive in NIIs, due to the limited competition in generation ([9,16,168]).

In this context, centrally managed storage stations lack opportunities for revenue stacking from various markets, which would add up to enable the investment's financial viability. Overall, the economic viability of storage investments is still questionable because of insufficient clarity and quanti-fication of the respective reward mechanisms. In the absence of such a straightforward compensation scheme, several ideas have been proposed in the literature to fill this gap, the most common of which are presented below:

- **The levelized cost of storage (LCOS)** concept ([80,160,169]). The LCOS, calculated by (1), can provide insight into the price a storage investment should receive per MWh of produced energy to constitute a viable investment. The LCOS definition is conceptually analogous to that of LCOE[4]. In (1), *capex* refers to the capital expenditure of the investment, *opex* denotes the fixed operational expenses, $C^{ess\text{-}charge}$ the cost of storage charging, $i$ the discount rate, $E^{ess}$ the annual energy storage discharges, and $y$ the evaluation period, in years. The rationale behind adopting this measure is that a tariff equal to LCOS will be defined for the energy injected into the grid by storage stations, a policy in principle similar to the FiT scheme for RES. However, if storage investments are not effectively coupled with renewables under a common portfolio, as is the case for HPS, the LCOS may prove inefficient, especially for fast response storages providing active power reserves, rather than energy arbitrage, reaching values exceeding 1000 €/MWh, ([169]).

  Note that the LCOS concept could be the basis for more complex remuneration schemes. More specifically, the Authors in [170] suggest the establishment of a pre-calculated tariff for each storage technology calculated with the LCOS metric. An additional time-varying subsidy is also foreseen to incentivize the storage devices to inject energy into the grid during high-price intervals and to absorb energy from the grid when prices are low or renewable curtailments occur. Such a proposal attempts to artificially recreate in NIIs the rules that intrinsically apply in fully competitive electricity markets of larger continental grids.

$$LCOS = \left( capex^{ess} + \sum_{y} \frac{C_y^{ess-charge} + opex_y}{(1+i)^y} \right) \Bigg/ \sum_{y} \left( \frac{E_y^{ess}}{(1+i)^y} \right) \quad (1)$$

- **The levelized cost of reserves (LCOR)** concept ([80,169]). The LCOR has been proposed in [80] to evaluate the necessary pricing of the primary up reserves allocated to storage facilities so that the entire station is economically viable. LCOR monetizes the value of the fast-response reserves provided to the power system by the storage station, remaining applicable only to reserve-providing units, such as the BESS, and ignoring any energy-oriented services that storage might concurrently provide (arbitrage, capacity adequacy, etc.). Hence, the LCOR approach is primarily valid for storages of lower energy-to-power ratios, installed to islands suffering from fast response services. LCOR can be calculated using equation (2), which disperses total investment costs over the annual allocated reserves to storage, $R^{ess}$, instead of energy discharges as in equation (1).

$$LCOR = \left( capex^{ess} + \sum_{y=1} \frac{opex_y}{(1+i)^y} \right) \Bigg/ \sum_{y} \left( \frac{R_y^{ess}}{(1+i)^y} \right) \quad (2)$$

[4] Both LCOS and LCOE metrics are acceptable for referring to storage stations, as they are both observed in the relevant literature.

- **The avoided system cost (ASC)** concept ([36,78,81,158,171]). This remuneration variant is based on the hypothesis that storage should be annually compensated according to the variable system cost (*VC*) savings attributed to its presence in the island grid, as in (3).

$$ASC = VC^{\text{w/o storage}} - VC^{\text{w/ storage}} \quad (3)$$

This concept relies on the establishment of a mechanism that would reward the storage with the difference between system operating costs with and without the station and is similar to the remuneration principles initially foresaw for the HPS plants (see Section 3.3). The main shortcomings of this concept can be identified as follows:

  - Remuneration level is strongly related to systems variable operating costs, which in turn are exposed to the variations of several uncertainty factors, such as the oil and $CO_2$ emissions rights prices. Thus, a continuing annual adjustment of stations' remuneration is required, according to the prevailing cost conditions.
  - An accurate and transparent quantification of systems operating costs without storage is needed for the pro per calculation of avoided system costs.
  - The dependence of storage income on the avoided system costs does not necessarily reflect the actual revenues required for the station to be financially viable. The avoided system cost concept alone, without the establishment of any complementing regulation mechanism, either exposes the station to lower revenues than the required ones to ensure feasibility, if the oil and/or $CO_2$ emissions rights prices fall below a certain level or, otherwise, might leave the investment with windfall profits.

- **Capacity-oriented remuneration** concepts ([160,172]). Capacity remuneration mechanisms (CRM) constitute an effective instrument to fill the missing money gap of system assets participating in electricity markets for the provision of capacity-related services, such as flexibility and contribution to resource adequacy. CRMs add value to the power capacity of specific assets for covering system needs, remunerating them through a fee (*CR*) computed by (4). The capacity remuneration is expressed in monetary units (€ or $) per MW of storage capacity credit ($P_{cc}^{ess}$) per annum. For island systems, the establishment of CRMs for the compensation of centrally managed storage facilities could be a solution to enhance storage income, provided there are additional revenue streams from electricity markets or other compensation schemes to complement storage required revenue for viability. In this sense, CRMs might find application to larger NII systems, which could accommodate organized markets, as in [160].

$$CR_y = P_{cc}^{ess} \cdot CP_y \quad (4)$$

- **Cost-of-service-oriented** (CoS) remuneration concepts ([1,21,63,173,174]). The underlying assumption for the implementation of a CoS-oriented remuneration policy is that the storage is entitled to receive an annual flat rate that would allow for station's viability. The annual flat fee should be such to cover the annualized capital expenditure and operating and maintenance costs of the storage station, as in (5), including any energy capacity augmentation required due to capacity fade. The CoS scheme should come along with the obligatory participation of the central storage station to all system services identified as important by the SO (balancing services, energy arbitrage, capacity adequacy, flexibility) so that the island to harvest the maximum benefit from the storage. This hypothesis is also in line with the centrally managed storage typology, where the SO directly manages and supervises its operation. The CoS concept can be effortlessly established in small and medium island power systems lacking organized electricity markets, as it remains similar in principle to the prevailing remuneration scheme for thermal generators, whose annual fixed and variable costs are fully covered.

$$CoS = capex^{ess} \cdot \left( \frac{i}{1-\left(1+i\right)^{-y}} + opex \right) \tag{5}$$

Table 3 summarizes the storage *capex* assumptions and the values of the aforementioned compensation schemes for indicative study cases investigating centrally managed storage facilities. A salient conclusion from Table 3 is that the avoided system cost approach dominates the existing literature. This observation is reasonable, given that the avoided system variable cost attributed to the presence of storage constitutes a realistic basis for evaluating the feasibility of the respective investment. However, the real-world application of this index to remunerate storage is strongly associated with the drawbacks of the ASC concept mentioned earlier. CoS principles also find application in several studies, especially those evaluating the capacity expansion planning of islands ([1,21,63,173,174]), where the annualized cost of each technology is an integral component of the problem formulation. Nevertheless, even in these cases, which mainly evaluate the pathways towards deep decarbonization, the CoS concept is not deeply analyzed as a tool to remunerate storage; rather, it is perceived as the means to calculate system costs accurately.

The LCOS concept is evaluated less in the literature, with its values presenting extreme variations, between 80 and 2600 €/MWh, strongly related to the specificities of the storage application and the respective *capex* assumption of the storage technology. LCOR and CRM schemes face the lowest recognition in the literature. On the one hand, LCOR is addressed to storage stations primarily providing reserve services, which have encountered limited applications so far. On the other hand, CRMs are directed to large island systems with organized markets, a situation out of the norm for most NIIs. Another important observation from Table 3 is that index values might vary significantly between case studies and storage technologies. This is quite reasonable given that each study case island presents different techno-economic characteristics, and each study makes different cost assumptions regarding storage investments.

*Table 3: CAPEX and values of economic evaluation indices for indicative studies evaluating centrally managed storage stations.*

| **Ref.** | *Capex BESS power* [€/kW] | *Capex PHS power* [€/kW] | *Capex BESS energy* [€/kWh] | *Capex PHS energy* [€/kWh] | **LCOS** [€/MWh] | **LCOR** [€/MWh] | **ASC*** [€/kW] | **CRM*** [€/kW] | **CoS* BESS** [€/kW] | **CoS* PHS** [€/kW] |
|---|---|---|---|---|---|---|---|---|---|---|
| [1] | 400 | 440 - 2870 | 150 | 5 | - | - | - | - | 96 | 52 - 312 |
| [21] | 304 - 500 | - | 89 - 154 | - | - | - | - | - | 89 - 149 | - |
| [36] | - | - | 700 - 2600 | - | - | - | 607 | - | - | - |
| [39] | - | *n/a* | - | *n/a* | - | - | 174 - 177 | - | - | - |
| [63] | *n/a* | - | 736 | - | - | - | - | - | 93 | - |
| [78] | - | - | - | - | - | - | 232 - 327 | - | - | - |
| [80] | 500 | - | 500 | - | 87 - 246 | 30-35 | 100 | - | - | - |
| [81] | *n/a* | - | *n/a* | - | - | - | 40 - 400 | - | - | - |
| [99] | 400 | - | 250 | - | 80 - 200 | - | - | - | - | - |
| [101] | *n/a* | - | *n/a* | - | - | - | 100 - 277 | - | - | - |
| [158] | 250 | 1155 | 140 | - | - | - | 35 - 118 | - | - | - |
| [160] | 400 | - | 200 | - | 106 | - | 176 | 114 | - | - |
| [169] | 400 | - | 250 | - | 1000 - 2600 | 14-30 | - | - | - | - |
| [171] | *n/a* | - | 261 - 287 | - | - | - | 376 - 719 | - | - | - |
| [173] | 1200 | - | *n/a* | - | - | - | - | - | 153 | - |
| [174] | 700 | 1200 | *n/a* | *n/a* | - | - | - | - | 102 | 133 |

** Not directly specified within the paper. Own calculations based on data available in each paper.*

## 4.3. Summary of centrally managed storage applications

Centrally managed storage facilities in island power systems dominate the relevant literature. Table 4 includes the papers dealing with the centrally managed storage concept. Table B2 of Appendix B and Figure 7 present additional details for the most representative ones. Note that except for the papers explicitly referred to in Table 2, which investigate storage within the HPS paradigm, the majority of the remaining literature examines storage as a standalone, centrally managed asset. From the available pool of papers addressing the centrally managed storage concept, the analysis primarily focuses on the most recent ones. Notably, Figure 7(b) shows that over 70% of the papers enumerated in Table 4 have been published within the last 5 ½ years.

From Figure 7(a), it is apparent that the literature on centralized energy storage stations explores island systems worldwide without being focused on a specific study case or country, unlike the HPS paradigm. Figure 7(c) specifies the storage technology investigated in the most representative papers, showing that batteries are the most favorable solution, explored in 61% of the reviewed applications. Only 18% of cases involve the analysis of stand-alone pumped hydro storage stations, a percentage substantially lower than that for the HPS paradigm. This indicates that centrally managed applications tend to prioritize shorter-duration storages that can be implemented without the need for specific geomorphological conditions.

Surprisingly, hydrogen storage holds a considerable share (13%) of the technologies evaluated for centrally managed applications, despite its current low technological readiness and maturity, preventing immediate implementation. This is attributed to the expectation of future technology development and a decrease in investment costs, enabling the massive implementation of long-duration and seasonally managed green-hydrogen storage facilities in islands to achieve fossil-fuel independence targets.

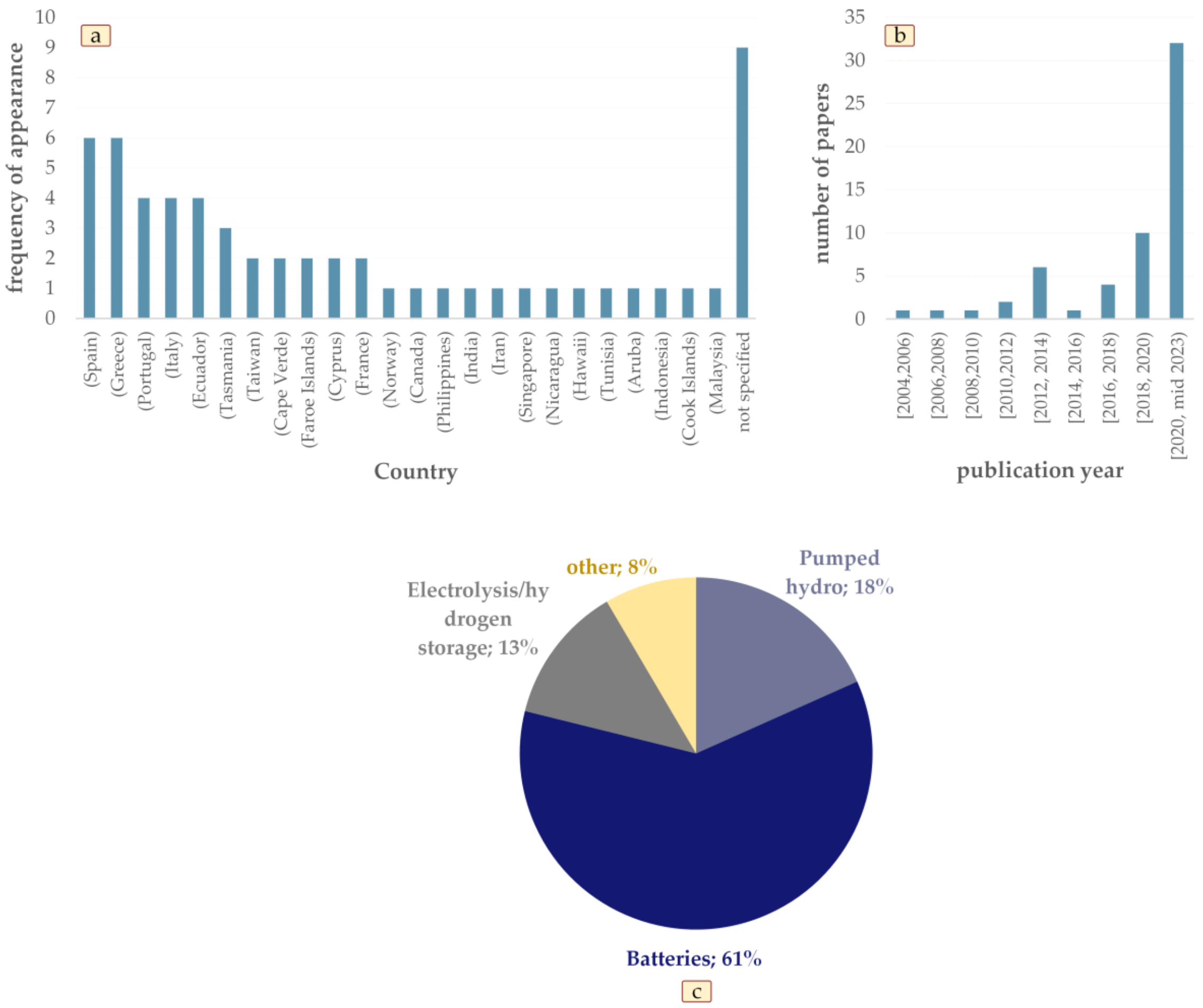

*Figure 7: (a) Country, (b) publication year and (c) storage technology investigated for the centrally managed storage concept papers reviewed.*

Table 4 reveals four topics of amplified interest for centralized storages: (a) the operation under increased RES penetration levels (29 papers), (b) the formulation of optimal expansion planning models and the consequent determination of the optimal generation mix for islands (22 papers), (c) the description of accurate generation scheduling processes that account for the dispatch of storage assets (18 papers), and (d) the determination of the optimal sizing of storage assets (16 papers). These topics are complementary to each other, as the pathways towards increased RES penetration levels are strongly related to both the optimal capacity expansion planning of a power system and the realistic, short-term scheduling of its assets in the day-ahead and intra-day markets.

Significant research has also been conducted on the dynamic behavior of island systems in the presence of storage and the feasibility of storage investments. On the other hand, the contribution of storage to resource adequacy in islands has received limited investigation, presenting opportunities for further research in this area. Note that this field of study still remains open for larger power systems experiencing high-RES penetration levels.

Other topics where central storage facilities are considered, though not as the key parameter, are related to the coupling of energy sectors. Additionally, some research has also been performed to compare the competition between storage and diesel units employing low-loading operation capa-bilities to achieve 100% RES penetration levels in small islands. Specifically, the research team of [60,175,176] argues that the small island systems can operate effectively under high RES penetration levels either by deploying battery energy storages to alleviate RES variations or by imposing the diesel generators to operate below their technical minimum loading levels, down to zero, to perform the same task.

*Table 4: Aim & Scope category per publication dealing with centrally managed storage facilities.*

| ***research topic*** | **Ref.** | **No.** |
|---|---|---|
| System dynamics & Frequency stability | [71][177][178][165][175][166][167][124][179][180] | 10 |
| Capacity adequacy & Capacity credit | [100][101][158] | 3 |
| Generation scheduling | [36][37][39][78][80][90][158][179][181][182][183][184][185][171][186][187][188] | 17 |
| Increase RES penetration levels | [1][21][43][45][47][49][50][51][52][54][55][59][60][62][63][65][66][81][99][158][178][186][187][188][189][190][173][174][191] | 29 |
| Optimal storage sizing | [21][36][50][51][80][81][99][158][171][186][188][189][173][174][191][192] | 16 |
| Investment feasibility | [78][80][99][189][191][192] | 6 |
| Capacity expansion planning | [1][21][44][45][47][49][51][54][55][59][60][62][63][64][65][66][189][190][173][174][191][193] | 22 |
| Sector coupling | [13][37][47][191] | 4 |
| Low loading operation | [60][175][176] | 3 |
| **Total *(single papers)*** | | 58 |

# 5. Conclusions and future work directions

## 5.1. Main findings

This paper has conducted a literature review to explore the various typologies, services, and applications of electricity storage and its role in attaining deep decarbonization targets in islands. The review encompasses over 190 scientific papers and institutional documents (laws, reports), focusing on those published within the last decade.

The analysis highlights that storage is a prerequisite in achieving renewable penetration rates exceeding 50%, providing system flexibility and shifting renewable energy to intervals of limited RES availability. For lower renewable penetrations, storage presence benefits the power system without, however, being instrumental in its operation. Additionally, it became clear that the pursuit of almost complete decarbonization (RES penetration over 90%) is possible; however, it requires a significant increase in renewable and storage capacities, which could be as high as six to seven times the system's peak demand. Nevertheless, the feasibility of ambitious RES penetration targets remains a topic of debate within the research community. Some researchers have expressed valid concerns about whether real-world island systems can adapt to this concept and the potential impact of such actions on the cost of electrification.

The review process identified three main storage typologies suitable for deployment in island systems: (a) storage coupled with RES within a hybrid power station, (b) centrally managed standalone storage installations, and (c) behind-the-meter storage installations. Of particular interest are the former two, which dominate the relevant literature. Hybrid power plants are of significant interest in geographically limited areas, mainly in the Mediterranean European countries, where frameworks have been established governing their market participation and remuneration. On the con-trary, centrally managed storages are present in islands all over the globe, as their management principles are simple and align with the vertically organized electricity markets thriving in isolated power systems. The behind-the-meter storage concept has proven generic and not intensively covered specifically for island systems so far in the existing literature.

The analysis of the relevant papers showed that the hybrid power plant concept mainly incorporates wind-powered hydro-pumped storage stations, aiming to mitigate the inherent RES volatility by controllably injecting energy into the grid via storage. Batteries, hydrogen storage, and CAES also exhibit a limited interest as storage technologies for HPS, especially in cases where the hydroelectric potential of the island has been depleted. The embracing of the HPS design by the independent producers and the particular interest in investing in such projects is associated with the incentives given for its development, reflected in well-defined compensation schemes. Thus, HPSs find several real-world applications in the Greek islands and the island of El Hierro in Spain, which, however, document technical limitations and shortcomings regarding the design fundamentals of the concept, especially in the synchronization between renewable generation and storage units during real-time operation.

On the other hand, centrally managed storage installations, which rely on a simpler management concept than HPSs, have been extensively examined in the literature, focusing on fast response storages, such as batteries. Literature review shows that central storage is associated with enhancing RES penetration levels and improving system dynamic behavior and frequency stability. However, despite the benefits associated with this storage concept, the financial viability of the relevant investments cannot be guaranteed to the extent that the mechanisms to remunerate their services are either absent or not clearly defined in the prevailing environment of island electricity markets.

## 5.2. Future work directions

Upon analyzing this review paper, it has come to light that several aspects related to storage in island systems require further consideration and research. Among them, the most significant ones are:

- **Establishment of well-defined remuneration schemes for centrally managed storage stations.** The literature review revealed that despite the benefits derived to the system by the centrally managed storages, there is no coherent framework regarding the remuneration of the respective storage investments for the services they provide to the grid. Although several proposals have been made, including the remuneration of energy-oriented services, reserves provision, contribution to resource adequacy, etc., the question regarding the financial viability of standalone storages in islands remains unanswered, as none of these metrics alone can capture the manifoldness of services provided by storage and suitably attribute a fair value to them. One significant parameter hindering the prospects of central storages is the lack of organized markets in small and medium island systems, which, in principle, would create multiple revenue streams and give a possible solution to the storage financial viability stalemate. However, the introduction of complex electricity markets structure in such systems does not seem to be valid, given the absence of competitiveness on the generation side. Thus, additional research is necessary to determine appropriate compensation schemes that are easy to implement and would provide potential standalone storage investors with a clear and straightforward understanding of the feasi-bility of their investments. In this context, the required legislative and regulatory interventions should also be identified, and proposals should be made to preserve that in the intensively regulated electricity market environment of islands, any storage investment decisions would not burden the electricity cost to be covered by the end consumer.

- **Exploring alternative storage architectures emphasizing the coupling of RES and storage.** The review process has highlighted two dominating storage concepts in island systems: the HPS design and the centrally managed storage. Storage collocated with renewable installations as a behind-the-meter facility, either to improve the dispatchability of variable RES generation or to act as an instrument to mitigate RES curtailments, has not been deeply investigated in the island's literature, particularly in terms of operating principles, market participation rules, sizing of the relevant storage-renewable components and remuneration strategies.
- **Investigating pathways towards full decarbonization of islands.** Several studies have examined the long-run optimal capacity expansion planning of island systems to attain their independence from fossil fuels and mitigate carbon emissions, acknowledging the role of renewables and storage to this end. However, in order to move toward such groundbreaking objectives and realistically achieve full decarbonization targets, a roadmap for the gradual implementation of storage and renewable projects should be established, specifying the operating policies, economic incentives, and legal actions required for a smooth transition. In the same context, research should also focus on coupling several sectors in islands, including electricity, heating, cooling, transportation, and desalination, with storage serving as the instrumental means linking renewable generation with the final consumption of each sector.
- **Identifying the capacity value of storage assets.** This issue is of utmost importance for isolated and interconnected power systems seeking high-RES penetrations and gradually abolishing their carbon-intensive thermal fleet. In such situations, the systems rely heavily on variable RES production and storage to meet their electricity requirements. However, this might lead to security-of-supply concerns due to the intermittency of renewable generation, particularly during periods of limited availability. In this sense, storage is possibly the only reliable enough system asset to address RES intermittency if appropriately sized and operated. Nonetheless, due to the intrinsic energy limitations of storage stations, estimating their contribution to resource adequacy is challenging, affected by several factors, such as the load profile, the storage architecture, sizing, technology, etc., and has not been thoroughly researched in the literature. Specifically for islands, which are vulnerable to disturbances and capacity adequacy risks, the capacity value of storage stations, either centrally managed or HPS plants, should be further investigated to quantify the levels of storage power and energy capacity that should be installed for each NII to operate securely in the absence of conventional generation.

## CRediT authorship contribution statement

**G. N. Psarros:** Conceptualization, Resources, Methodology, Software, Validation, Formal analysis, Investigation, Data curation, Writing - original draft, Visualization. **P. A. Dratsas:** Formal analysis, Investigation, Writing - original draft. **S. A. Papathanassiou:** Supervision.

# Declaration of competing interest

The authors declare that they have no known competing financial interests or personal relationships that could have appeared to influence the work reported in this paper.

# Appendix A: Review methodology and statistics of the documents analyzed

To conduct this study, we initially screened over 400 scientific documents, which included research and review articles, conference papers, institutional documents, and books related to island power systems. Subsequently, we reviewed nearly 300 of them. Out of these, 195 documents were selected for inclusion in our work based on the following criteria:

- They exhibit higher research quality, in the sense that they furnish valuable results and methodologies that are of significant interest, contributing to useful insights, generalized conclusions, and an understanding of the state-of-the-art.
- They emphasize the introduction of storage in islands, pathways to achieving high or very high RES penetration levels, highlight the crucial role of storage, present practical, viable and applicable alternatives for storage remuneration in islands, identify services storage can offer to island systems, etc.
- They have been published in the last 10-15 years, offering up-to-date information on the field. A few older publications (from 2004 onward) have also been included to illustrate the evolution of the state-of-the-art over time.
- Articles of archival value were reviewed and cited where necessary throughout the paper to support arguments.

During the initial screening, articles that focused on storage without explicitly examining storage in island contexts or the transition of islands to fossil-fuel independency were excluded, except for a few review articles in the Introduction and throughout the paper that were necessary to support specific statements. All the scientific articles selected for inclusion in the manuscript are indexed in databases such as *scopus.com* and *scholar.google.com* and retrieved directly from the official websites of their publishers.

As Figure A1 shows, out of the 195 documents included in the paper, approximately 83% are research (68%) and review (15%) articles published by reputable publishing houses in the field, including Elsevier (50%), IEEE (15%), MDPI (12%), and IET (4%). About 10% of the literature consists of conference papers, primarily supported by IEEE and IET, while the remaining 7% includes books, websites, and institutional documents such as laws, European Commission decisions, and reports. The frequency of appearance of each journal title in the reference list is depicted in Figure A2. Figure A3 categorizes the reviewed papers into publication periods using a 2-year interval for those published up to 2020, and groups them all together for subsequent years. Notably, 75% of the papers included in our work were published within the last 10 years, from 2014 to mid-2023.

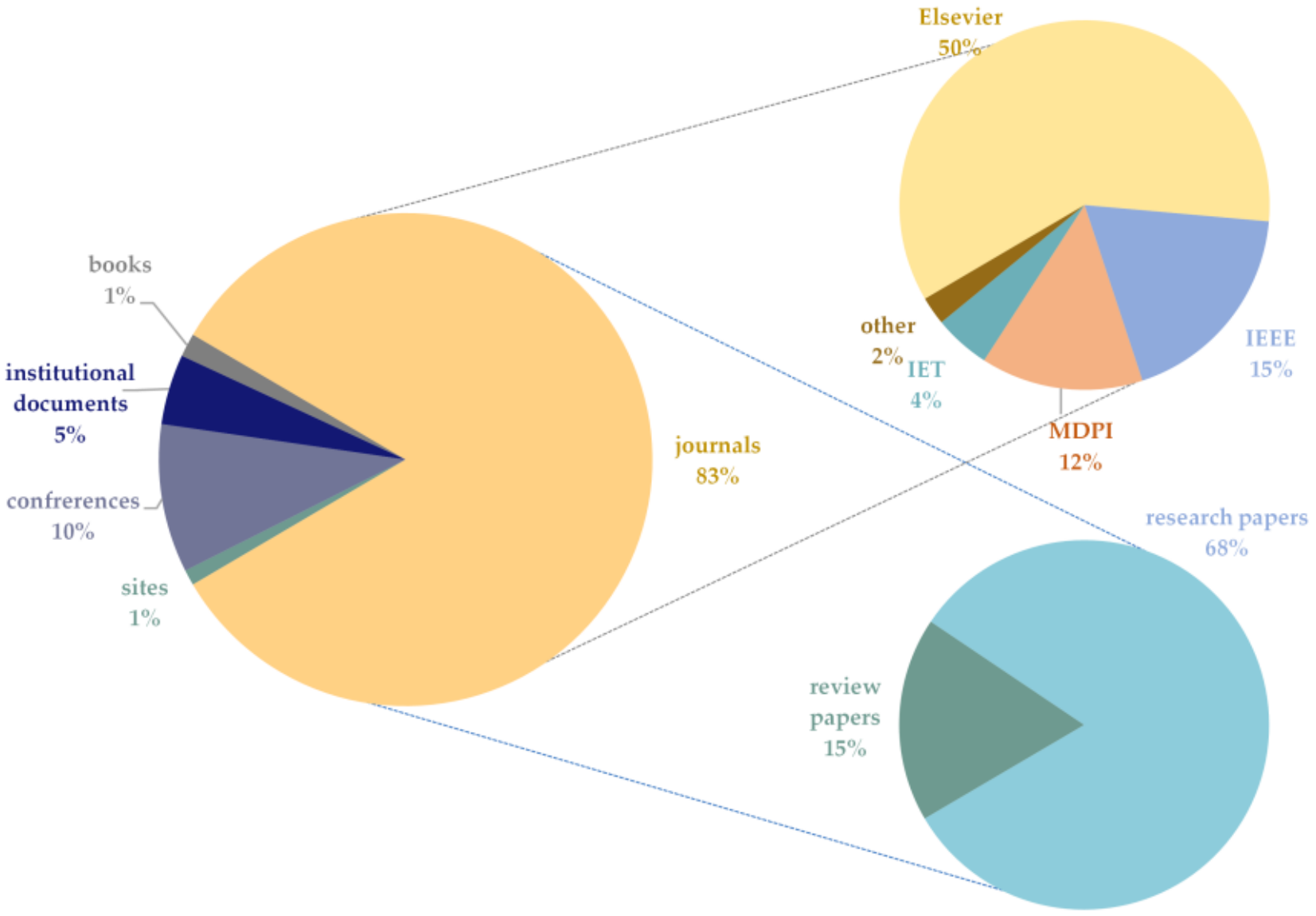

*Figure A1: Classification of the reviewed documents.*

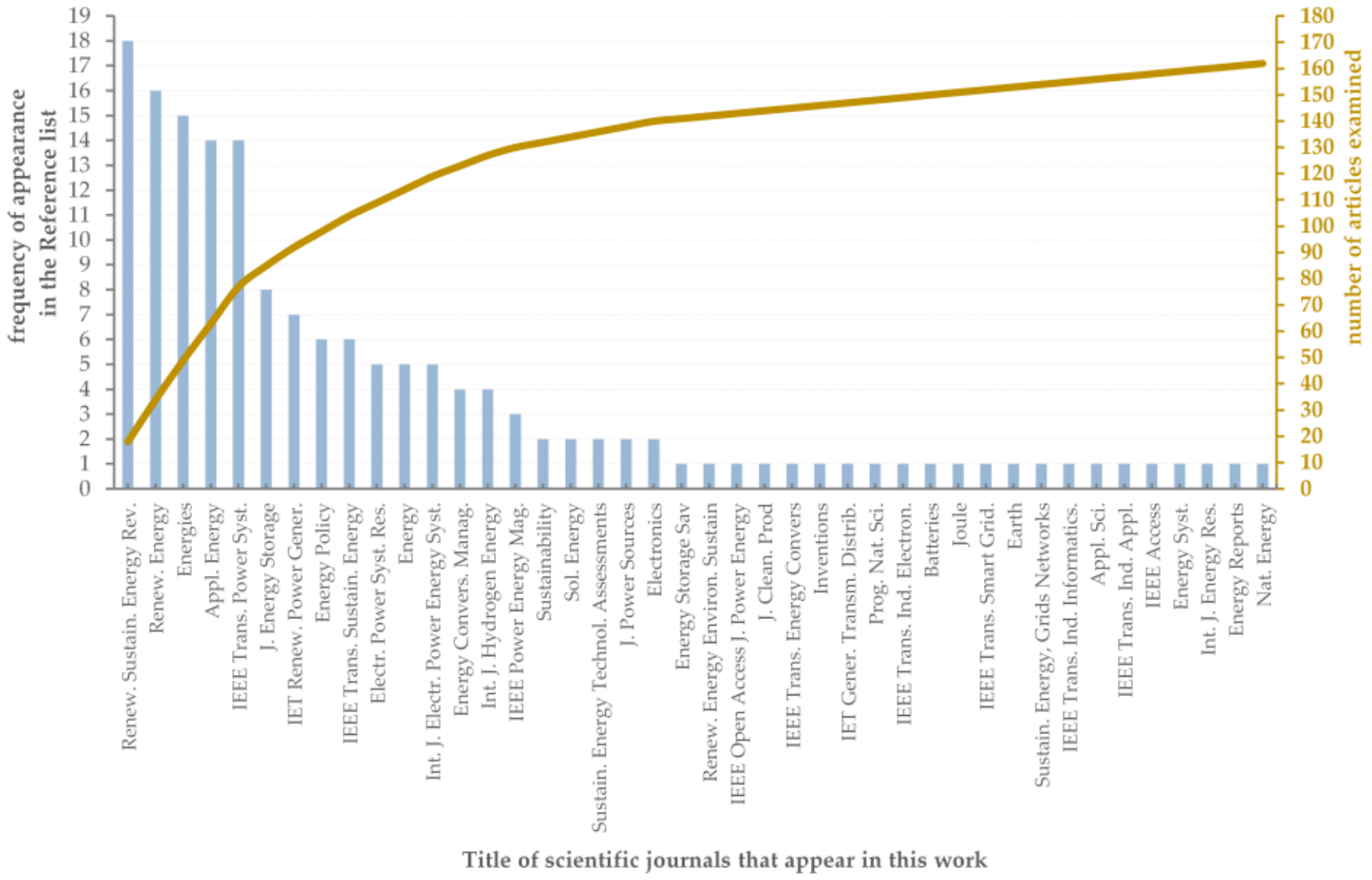

*Figure A2: Frequency of appearance of each journal title in the reference list.*

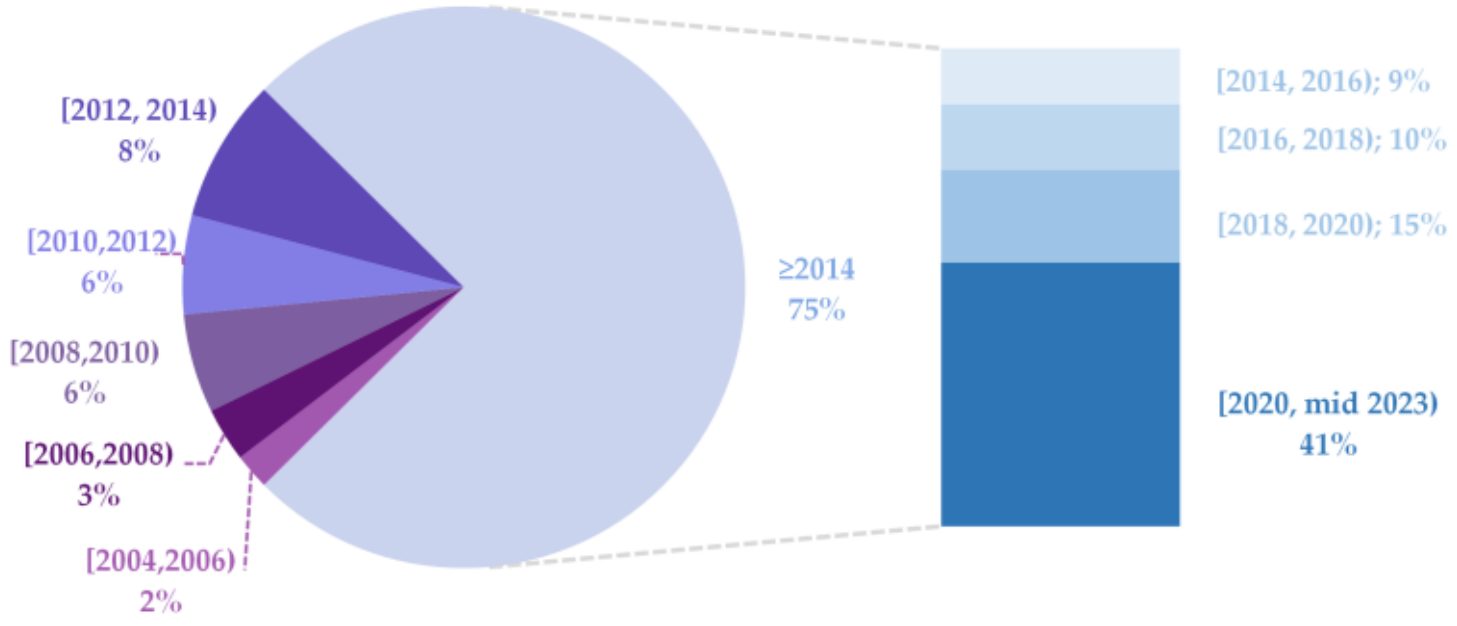

*Figure A3: Number of documents reviewed in this study per year of publication as a percentage of all documents*

# Appendix B: Details of the investigated storage studies

## Available HPS studies

*Table B1: Storage and RES technology of HPSs investigated in the literature, island system and study objective.*

| **Ref.** | *year* | *RES technology* | *Storage technology* | *Island system* | *Aims & scope* | *Main findings* | *Comments & critical points* |
|---|---|---|---|---|---|---|---|
| [42] | 2005 | WFs | Pumped hydro | El Hierro (*Spain*) | • Proposing suitable models for the sizing of hybrid power plants components and the optimal operation of the island, examining several operating policies (6 in total) | • The mathematical formulation of the 6 strategies is provided. The proposed model can be applied to analyze the characteristics and costs of technically feasible and commercially available combinations. | • The paper introduces the concept of HPS consisting of WFs and PHS operating in a NII system, without implementing the proposed method on a specific case study. This follows in [48].<br>• The proposed strategies are developed through a heuristic approach. |
| [48] | 2005 | WFs | Pumped hydro | El Hierro (*Spain*) | • Applying the models developed in [42] to conclude on the size of hybrid plant and to quantify system benefits and costs for each of the 6 proposed operating policies. | • Optimal results are achieved when the applied strategies allow flexible wind energy utilization.<br>• The economically favorable HPS configuration involves pumps of various capacities and turbines with amplified power rating.<br>• PHS systems with low energy-to-power ratios are optimal from a system cost perspective. | • Since the study was conducted in 2005, all technical and economic assumptions could be updated. |
| [119] | 2007 | WFs | Pumped hydro | Crete, Serifos, Lesvos (*Greece*) | • Investigating the possibilities of integrating HPS in island systems to reduce system costs, increase the renewable penetration levels and enhance the capacity credit of renewable stations via the dispatchability provided by the pumped hydro-storage plants | • The generation cost in islands is more predictable with the introduction of HPS, as it relies less on the oil price variations.<br>• The feasibility of HPS remains sound even when more strict limitations are imposed to the station regarding the levels of direct wind penetration. | • The hydro-turbine is set into operation when the demand surpasses a predetermined level.<br>• Operational decisions are made on an hourly basis using a heuristic approach. |
| [121] | 2008 | WFs | Pumped hydro | Crete (*Greece*) | • Propose a methodology to sizing the pumped hydro storage components, including, turbines, pumps, reservoirs and the penstock characteristics.<br>• Investigate the financial viability of HPS stations via developing a suitable economic model. | • The economic viability of the HPS project depends strongly on public funds' contribution to initial capital costs and the electricity price difference between purchasing and selling energy to the local grid.<br>• Financial prospects improve with a higher installed power of the hydro turbine; however, adverse results may arise for turbines above 3–4 MW. | • The annual wind power curtailments are used as input parameters to optimize the size of the HPS.<br>• The simulation of HPS operation lack detailed description. |
| [120] | 2009 | WFs | Pumped hydro | 3 islands (*Greece*) | • Proposing suitable operating policies for the integration of HPS stations to island systems in order to increase the RES penetration levels of the islands, minimize system costs and ensure the feasi- | • HPSs significantly enhance WFs penetration and offer firm capacity, substituting expensive peak units.<br>• HPSs prove to be attractive investments under the examined pricing framework, without excessively burdening the operat- | • A simplified peak shaving approach is adopted for the calculation of capacity credit. |
| | | | | | bility of the hybrid plants investments. | ing cost of island systems. | |
| [98] | 2010 | WFs | Pumped hydro | Crete, Lesvos, Serifos (*Greece*) | • Examining system reliability in the presence of hybrid power plants, using the LOLE and ENS reliability metrics.<br>• Calculating the capacity credit of the combined WFs-pumped hydro storage station utilizing the ELCC methodology.<br>• Assess the prospects of wind energy with or without pumped storage systems to substitute the installation of conventional power plants. | • Integration of PHS improves the capacity credit (CC).<br>• CC decreases as the size of PHS increases.<br>• Over-sizing wind installed capacity to provide additional energy for turbining is a reasonable energy planning strategy. | • The convolution method is used to assess the system resource adequacy.<br>• The same typical wind data series has been used in all three examined island systems. |
| [41] | 2010 | WFs | Pumped hydro | Ikaria (*Greece*) | • Proposing operating policies for the operation of a complex HPS (built upon the fundamentals of [120]) and the island as a whole, targeting to the smooth integration of the project.<br>• Quantifying system benefits deriving from the introduction of the said HPS to the island system. | • The HPS significantly increases RES penetration, potentially exceeding 50% per annum.<br>• Efficient utilization of wind and hydro potential by the HPS allows offering firm capacity to the NII system and substituting conventional peaking units. | • A time-series-based simulation model is used, reflecting the power equilibrium and ignoring phenomena.<br>• Reserve requirements are ignored. |
| [147] | 2010 | WFs & PVs | batteries | n.s. | • Review the state of simulation, optimization and control technologies for the stand-alone hybrid solar–wind energy systems with battery storage. | • A comprehensive RES resource analysis is crucial in the initial sizing of an HPS.<br>• The integration of artificial intelligence into the energy management system is promising for further reducing the system's cost in the future. | • Analysis focuses on the meteorological data used in feasibility studies of the HPS, the modeling of its components, the criteria used in relevant works for HPS sizing optimization, and the optimization methods applied. |
| [154] | 2010 | WFs | Pumped hydro | Lesvos (*Greece*) | • Propose a methodology to determine the sizing of HPS components and concurrently estimate the wind power penetration hosting capacity of an island. | • For PHS to store a greater portion of the excess wind energy, there is a need not only for a large upper reservoir but also for high turbines' nominal power. | • An hourly approach for HPS operation is adopted, using as input wind power rejections. |
| [155] | 2010 | WFs | Pumped hydro | Lesvos (*Greece*) | • Determine the sizing of HPS constituents and evaluate the economic viability of project implementation via maximizing the net present value index. | • The optimum HPS configuration corresponds to RES penetration levels of 19%. Other suboptimum but feasible solutions can lead to higher RES penetration levels.<br>• The price of the guaranteed electricity sold to the local utility network is a critical parameter for the investment's feasibility. | • The energy model presented in [154] is used.<br>• NPV and payback period are evaluated for the economic viability assessment. |
| [32] | 2010 | WFs | Pumped hydro | 23 islands (*Greece*) | • Examining the potential of introducing the wind-pumped-hydro HPS concept in 23 islands of Greece<br>• Performing a pre-feasibility study regarding the economic viability of the wind-pumped-hydro HPS installations in the examined islands, by comparing their investment costs with the avoided fuel costs of thermal units | • There is a considerable market in Greece for HPS comprising WFs and PHS.<br>• For three representative levels of HPS penetration (30%, 50%, 70% of hydro-turbine's peak demand supply), the renewable energy contribution reaches 21%, 43%, 64%, requiring investment costs of 927, 2297, and 4309 M€, respectively, and substituting conventional capacity of 356, 593, and 831 MW. | • A simplified method is applied to size HPS alternatives, considering system energy demand without performing a detailed analysis of HPS operation or its interaction with the power system. |
| [123] | 2012 | WFs | Pumped hydro | Karpathos–Kasos | • Determining the optimal sitting of the PHS station between three candidate locations (the lower reser- | • The proposed project is expected to have attractive economic indicators, regardless of the availability of a subsidy, provided | • The analysis focuses on HPS's components sizing.<br>• Curtailments of HPS's wind power due to system |
| | | | (*Greece*) | voir of the project will be the sea)<br>• Investigating projects economic viability for given tariffs remunerating both the energy services and the contribution to resource capacity provided by the HPS plant. | that the selling price of energy is set at 0.3552 €/kWh and the selling price of guaranteed power is 127 €/kW year (the mean specific annual energy production cost from the system is at 0.249 €/kW h in 2008).<br>• Under the mentioned conditions, the project's payback period is estimated at 5–6 years. | limitations are not considered.<br>• The individual costs of HPS components are provided while a long-term loan covering 50% of the total project's CAPEX is assumed for the analysis. |
| [142] | 2013 | WFs | Pumped hydro | Crete (*Greece*) | • Static (power flow) and dynamic (frequency-response) analysis has been performed in the presence of hybrid power stations and solar thermal power stations. | • Results reveal that from a power flow analysis perspective few reinforcements of the existing transmission system are required to accommodate HPS and solar thermal plants. | • Contingencies, including transmission line loss for static security analysis, unit tripping, and a three-phase fault for dynamic security examination, are taken into account. |
| [148] | 2014 | WFs | Pumped hydro | Crete (*Greece*) | • Proposing a generation scheduling optimization method for island systems incorporating HPS. | • HPS can substantially reduce total production costs increase RES penetration. | • Optimization model for day-ahead scheduling<br>• Annual simulation for sizing HPS components.<br>• System-level RES curtailments are ignored. |
| [126] | 2014 | WFs | Pumped hydro | Lesvos (*Greece*) | • Investigate the optimum sizing of PHS components (WFs, pumps, turbines, reservoirs) when examining the integration of the hybrid plant either from an independent power producer perspective (max profit objective), or from a system-level viewpoint (min-cost & max RES penetration). | • Energy and capacity tariffs significantly impact the optimal sizing of HPS components.<br>• The sizing of HPS components is moderately affected by the investment costs, with the CAPEX of WF being the most significant factor.<br>• The incorporation of a suitably sized HPS, even in systems with low thermal generation costs, can decrease the island system LCOE and enhance RES penetration. | • For the HPS operation the simulation models presented in [41] and [120] are used<br>• Genetic algorithms are applied as the optimization method, along with Pareto optimality concepts. |
| [141] | 2014 | WFs | Hydrogen or CAES | Karpathos (*Greece*) | • Review operating principles of hydrogen storage and CAES.<br>• Compare the operation of two hybrid stations involving wind farms and the respective storage technologies in terms of performance, efficiency and technology costs. | • Hydrogen storage has relatively high response time and decreasing efficiency over its lifetime.<br>• Two-stage CAES systems exhibit a roundtrip efficiency of approximately 32.5%, while single-stage systems have a higher efficiency of about 55.6%. CAES response time is notably lower than that of hydrogen storage. | • The paper does not delve into power system operation.<br>• Economic aspects for both technologies are provided, which may require updating. |
| [125] | 2015 | WFs | Pumped hydro | *not specified* (*Greece**) | • Investigating the dynamic behavior and stability of small-medium sized island in the presence of HPS<br>• Assessing the impact of different HPS operating policies on system stability and the effect of substituting thermal generation by HPS production on system dynamics | • Substituting diesel units with hydro turbines worsens system frequency response across all scenarios, particularly during severe events.<br>• The contribution of wind turbines to frequency regulation results in decreased frequency deviations.<br>• HPS with certain technical capabilities can exhibit a superior dynamic response compared to fast diesel units. | • Two severe disturbances are examined: an outage of a diesel unit and a bolted 3-phase fault on the transmission system under high wind conditions, when all WFs operate at nominal power.<br>• Frequency fluctuations in normal operation when wind speed surpasses 8 m/s are extremely intense compared to the smooth frequency pattern for wind speed values below that threshold. |
| [132] | 2016 | WFs & PVs | Batteries, pumped hydro, electrolysis | 7 islands (*Greece*) | • Optimal dimensioning of HPS components in 7 different islands which incorporate diversified characteristics (Agios Efstratios, Castellorizo, Karpathos–Kasos, Astypalaia, Rhodes, Samos, Crete)<br>• Evaluating the economic feasibility of each exam- | • The examined HPS are technically and economically viable, regardless of the insular system's size.<br>• The electricity prices may be reduced due to HPS presence.<br>• In small power systems, economic feasibility for HPS is attained for annual RES penetration levels over 90%. | • The HPS operation is determined on an hourly basis adopting a heuristic approach.<br>• Detailed investment cost assumptions for HPS are not included. |
| | | | | | ined HPS | | |
| [134] | 2017 | WFs | Pumped hydro | Crete (*Greece*) | • Proposing an optimal bidding strategy for HPS participating in the markets of Greek island systems.<br>• A stochastic bi-level optimization model is presented for determining the optimal energy offer/bid decisions of an HPS producer. | • A comparison with deterministic approaches for HPS indicates that the proposed model outperforms by formulating higher energy offers, leading to increased HPS profits without incurring high imbalance costs in actual operation. | • The proposed method considers uncertainty only in the power production of the WFs owned by the hybrid producer, with generation from WFs external to the hybrid pumping station treated as deterministically known.<br>• The method is demonstrated using an indicative time period.<br>• Model assumes that the HPS-O has perfect knowledge of system operating conditions. |
| [133] | 2017 | WFs | Pumped hydro | Crete (*Greece*) | • Proposing an optimal bidding strategy for the HPS to participate in the island market, with risk aversion techniques. Concept expanding the idea of [134]. | • As the HPS producer becomes more risk-averse, the expected profit for a given confidence level decreases.<br>• The expected profit decreases while the conditional value at risk increases, indicating a decrease in risk exposure with greater risk aversion. | • The comparison of computational complexity between the Mathematical Problem with Equilibrium Constraints and the MILP formulations possibly requires further research.<br>• An extended study, covering a longer period (e.g., a year), could offer further insights to analyze the impact of HPSs on NII systems. |
| [77] | 2018 | WFs | Pumped hydro | El Hierro (*Spain*) | • Evaluating control strategies for the contribution of variable speed wind turbines to frequency regulation. | • Wind turbines' inertial control can enhance system dynamics in all cases.<br>• Wind turbines' proportional control can improve frequency nadir values when a good wind regime exists. Otherwise, this control might compromise system stability during low wind production intervals. | • An approach based on exhaustive searches is proposed for the adjustment of the PHS and variable speed wind turbines controller gains. The benefits of the adjustments of the PHS and variable speed wind turbines used together or individually are analyzed. |
| [150] | 2019 | WFs | Pumped hydro | El Hierro (*Spain*) | • Determining the water value of stored energy in the pumped hydro storage station over a prolonged time horizon (2 weeks) for the accurate formation of the day-ahead scheduling problem of the island | • The day-ahead generation schedule obtained by the proposed model aligns closely with the one obtained from a detailed MILP model, revealing its effectiveness.<br>• The computational time required for the proposed model is significantly lower compared to the MILP model. | • A two-stage stochastic LP optimization model is developed considering demand uncertainty.<br>• The optimization horizon of the two stages is 14 days and 24 hours.<br>• A custom simulation tool is used, developed in GAMS |
| [143] | 2019 | WFs | Pumped hydro | El Hierro (*Spain*) | • Three different frequency control schemes are proposed to reduce the contribution of PHS turbines to the frequency control of the island during pumping mode (short-circuit operation), avoiding energy losses and thus increasing the RES penetration levels of the island. | • The proposed control schemes effectively address frequency regulation resulting from wind power fluctuations. Notable improvements are observed when wind turbines contribute to frequency regulation, while flywheels show less significant enhancements.<br>• In cases of sudden loss of generation, especially when operating without thermal units, pump shedding becomes necessary to keep the frequency within acceptable limits. | • Only the sudden disconnection of the wind generator with the highest power connected to the net is examined as abnormal operating condition. |
| [82] | 2020 | WFs & PVs | batteries | *not specified* (*Greece**) | • Proposing a UC-ED optimization method for the integration of HPS in small, isolated power systems. | • External to the HPS WFs benefit from reduced curtailments, while the security of operation is improved due to fast-response battery reserves. | • A detailed methodology is presented for simulating an isolated power system with HPSs. A UC-ED is formulated as a MILP problem with 24h/12h |
| | | | | | • Examine system operation and evaluating the economic viability of HPS investments.<br>• Compare the effectiveness of different HPS configurations to the achievable RES penetration levels and system costs reduction | • The HPS reduces system generation costs, assuming realistic oil prices above $70/bbl.<br>• A combined wind-PV-BESS HPS outperforms HPS with only WF or PV, offering higher RES penetration at a lower LCOE. | look-ahead horizon for day-ahead and intraday scheduling.<br>• Optimizing HPS component sizing, energy offering policy, and suitable remuneration schemes have not been evaluated. |
| [153] | 2020 | WFs | Pumped hydro | Crete (*Greece*) | • Identify and highlight the role of PHS station in the operation of the island system and its future interconnection with the mainland grid.<br>• Propose a detailed model for the pumped-hydro power plant and size its components | • PHS absorbing excess wind energy from WFs can result in competitive electricity prices.<br>• HPS implementation offers significant energy savings, ensuring grid stability, during periods of island grid isolation and after interconnection with the mainland. | • HPS operation is not simulated as an integral part of the island power system.<br>• Only WF curtailments and demand load timeseries are input for HPS sizing. |
| [144] | 2020 | WFs | Pumped hydro | El Hierro (*Spain*) | • Investigate the contribution of variable speed wind turbines to system dynamics, when providing inertial response and primary frequency control services, either jointly or individually. | • In normal operating conditions, WFs enhance system inertia and primary control, improving both upper and lower system frequency values.<br>• Wind turbines' contribution to frequency and inertial response reduces gate opening movements in hydroelectric units, extending their lifetime.<br>• In unexpected events, wind turbines contribute to frequency regulation, reducing load shedding events and stabilizing system frequency, even with action delays up to 500 ms. | • Analysis simulates realistic system events, including fluctuations in wind speed and the outage of the largest generation unit.<br>• 100% RES penetration scenarios are examined |
| [61] | 2020 | WFs | Batteries | Astypalaia (*Greece*) | • Examine several battery energy storage configurations to conclude on the optimal sizing of the hybrid power station that maximizes the net present value of the investment. | • Increasing BESS capacity decreases thermal units' production, and wind curtailments.<br>• Due to high initial battery costs, a larger BESS did not improve the HPS's financial performance.<br>• The fast response of the BESS improved system stability, enhancing robustness in potential disturbances after HPS installation. | • A rather simplified approach is adopted for HPS operation.<br>• The charging and discharging power capacity of BESS in not optimized. |
| [135] | 2020 | WFs | Batteries, pumped hydro | *not specified* | • Investigate the real time coordination of HPS assets so as the plant to avoid exposure to imbalance penalties.<br>• Proposing a real-time optimization method to maximize HPS profits from market participation | • Increasing the power capacity of HPS batteries brings significant benefits in terms of RES exploitation.<br>• Leveraging available tolerances for energy and power deviations from dispatch orders can considerably enhance plant revenues. | • The applied operating policy results in a low life expectancy for BESS, whether penalizing or not state-of-charge (SoC) variations (7.85 and 4.22 years, respectively). |
| [149] | 2021 | WFs | Pumped hydro | El Hierro (*Spain*) | • Determining the water value of stored energy in the pumped hydro storage station as in [150]<br>• Introducing the intra-hour variability of wind production in the formulation to highlight the hidden system inflexibilities and wind curtailments of the hourly time-step models | • The proposed model generates schedules capable of integrating more wind energy compared to models neglecting intra-hourly wind variability with a moderately higher scheduling cost.<br>• Utilizing the Benders decomposition technique for the solution approach reduces the computational burden of the model, making it suitable for day-to-day TSO tasks. | • A two-stage stochastic LP model was developed to assess the storage opportunity cost and a deterministic MILP-based UC model was used to estimate the next-day generating schedule. |
| [151] | 2021 | WFs | Batteries | Guadeloupe (*France*) | • Propose a model predictive control strategy for the scheduling of battery energy storage based on a | • Multiple case studies are examined, emphasizing the economic and technical aspects of HPS operation. The effectiveness of | • The effect of parameter tuning was shown through various co-simulation scenarios, enabling the strat- |
| | | | | | multi-objective management algorithm. | the proposed strategy, relying on Model Predictive Control, is demonstrated. | egy to be adapted over 28 days to assess its performance. |
| [46] | 2021 | WFs & PVs | Pumped hydro & batteries | El Hierro (*Spain*) | • Investigate the development of self-sufficient island system both in power and heat sector.<br>• Implement innovative technologies conclude on the most suitable for the study case island. | • Adopting a wind-power PHS system for electricity generation and distributed solar thermal fields for domestic hot water preparation demonstrates promising performance compared to traditional systems.<br>• PV solar fields, implemented to meet the increased electric energy demand due to population growth, demonstrated significant energy savings. | • The TRNSYS software utilized for system simulation lacks a module for simulating the Pelton turbine. Equations representing the Pelton turbine's working principles were implemented to assess power production from energy conversion. |
| [145] | 2021 | WFs | Pumped hydro | El Hierro (*Spain*) | • Investigate different control strategies for variable speed wind turbines, Pelton turbines and variable speed pumps to minimize the load shedding events after critical disturbances in the power system (loss of largest generator). | • Incorporating Pelton units as synchronous condensers reduces activation probabilities of load shedding.<br>• Variable Speed Wind Turbines contributing to frequency regulation improves the current under-frequency load shedding scheme, introducing, however, an oscillation phenomenon in the system frequency.<br>• Variable Speed Pumps are the most effective control strategy in improving the current under-frequency load shedding scheme. | • System updated scheduling after the clearance of the disturbances to restore reserves levels within the normal range is not evaluated. |
| [99] | 2021 | WFs | Batteries | *not specified* | • Compare the system benefit deriving from the introduction of HPS or centrally managed storage facilities in island systems and conclude on the most favorable storage scheme. | • In terms of achievable RES penetration HPS concept is competitive to centralized ESS, however it is a less cost-efficient option both for system and investors.<br>• Monetizing the value of capacity firming services makes some HPS configurations a viable option from system perspective. | • The development of new WFs is only investigated.<br>• The analysis considers only BESS, with no exploration of other storage technologies. |
| [152] | 2022 | Offshore WFs & CSP | Pumped hydro | Rhodes (*Greece*) | • Examine the RES technologies that have not yet been considered under the HPS concept in island systems.<br>• Investigate the optimum sizing of HPS aiming to achieve RES penetration levels higher than 70%.<br>• Assess the economic feasibility of HPS. | • The optimum configuration might achieve RES penetration levels of 79%, much higher from the initial 16.9%.<br>• Assuming a 25%/75% equity/bank loan, the payback period is assessed at 5.2 and 16 years for the WF/PHS and the CSP systems, respectively. | • The mathematical model is quite simplistic, not relying on optimization to simulate the power system and the HPS configurations. This might over/underestimate the energy results obtained for the HPS solutions envisaged. |
| [156] | 2022 | WFs | Pumped hydro & hydrogen | Leros (*Greece*) | • Investigate the optimum sizing of the HPS aiming to limit the exclusive use of fossil fuels for power generation.<br>• Explore the development of HPS both in electricity and transportation sector. | • A sensitivity analysis of investment feasibility is conducted by considering different electricity selling prices and calculating several fundamental economic indices. | • Hydrogen is exclusively used to supply a hydrogen ship. Interactions with the power sector could be introduced in an attempt to couple different sectors by utilizing hydrogen as the enabler.<br>• The financial assumptions about HPS components are not provided. |
| [157] | 2022 | WFs | Pumped hydro or hydrogen | Lemnos (*Greece*) | • Investigate the development of HPS with two operating options - storing wind energy as hydraulic energy or hydrogen. | • The water needs of the island are fully covered in both scenarios. PHS identified as the most reliable solution for energy needs.<br>• Hydrogen production could surpass the energy output of the hydroelectric plant depending on wind energy surplus. | • The paper's primary goal is to optimize water and energy needs.<br>• The simulation model oversimplifies system operation to cover energy, water supply, and irrigation needs. |
| [140] | 2023 | WFs or PVs | Pumped hydro or batteries or CAES | *modified islanded version* | • Propose a sensitivity-based three-phase volt/var control and energy management approach that produces the optimal settings of HPSs, step voltage regulators, capacitor banks and thermal generators to minimize the generation cost, for a 24-h period. | • The method successfully minimizes generation costs while satisfying various constraints, including the technical limits of diesel units, maximum direct penetration limits of RES, state of charge limits of ESS, and three-phase voltage limits of network buses.<br>• The proposed method demonstrates reasonable computation times. | • The approach assumes a centralized communication structure which is not applicable in distributed or decentralized architectures.<br>• The method ignores the uncertainties of loads and renewables. |
| [146] | 2023 | Wave energy & PVs | Batteries | *modified islanded version* | • Propose a novel source-storage coordination strategy using PVs, wave energy and BESS. The energy exchange model and an energy management scheme are formulated to keep balance between energy harvesting and battery life loss. | • The proposed source-storage coordination strategy effectively leverages the adjustment capability of wave energy generators and resolves the trade-off between energy harvesting and battery life loss.<br>• The strategy enhances the flexible adjustment performance of wave generators, leading to a significant improvement in the operational state of NII microgrids. | • An effective approach for island microgrids to incorporate impulsive components like wave energy generators is introduced.<br>• The analysis does not consider the microgrid's adjustment capability, resulting in a relatively conservative solution. |

* *Not directly specified within the paper. Assumption based on the data available.*

## Available centralized storage studies

*Table B2: Storage technology, island system, island RES technologies and study objective.*

| **Ref.** | *year* | *RES technology* | *Storage technology* | *Island system* | *Aims & scope* | *Main findings* | *Comments and critical points* |
|---|---|---|---|---|---|---|---|
| [50] | 2011 | WFs, PVs | batteries | Agios Efstratios (*Greece*) | • Addressing issues related to the required operating strategies to achieve deep decarbonization of small island systems, quantify system economic benefits, optimizing system assets' size and conclude on the feasibility of investments | • RES penetrations up to 60% are feasible and economically viable with the proper sizing of RES and storage capacities.<br>• Higher penetrations are still possible but result in increased costs, oversizing, and underutilization of RES potential. | • The study focuses only on lead-acid BESS technology with less favorable technical characteristics, such as a minimum SoC at around 45%. An updated should incorporate the latest advancements in battery technologies. |
| [71] | 2012 | WFs, PVs | ultracapacitor | Guadeloupe (*France*) | • Dynamic frequency control provided by storage to support islands stability after the loss of generation under high-RES penetrations.<br>• Analysis focuses on transient frequency phenomena. | • Adopting appropriate control strategies for modern distributed ESS, the impact of non-inertia generation like wind and PV on island system dynamics can be mitigated and potentially completely offset. | • The RES penetration levels examined remain relatively low, between 12% and 30%.<br>• Further investigation is required in aspects related to the integration of RES in isolated systems, such as RES variability and forecast accuracy. |
| [78] | 2013 | WFs | yes *(tecnology neutral)* | Gran Canaria, La Gomera (*Spain*) | • Assessing the impact on island system economics deriving from the introduction of storage stations performing energy arbitrage, providing frequency services or both. | • Combining spinning reserve provision and peak-shaving functionality for ESS, system savings are improved compared to providing only one service.<br>• In larger power systems, the presence of ESS is more advantageous. | • Only spinning reserves requirements are formulated.<br>• Analysis results rely upon indicative weekly optimization, missing critical information of RES and storage operation over the year. |
| [36] | 2016 | WFs | batteries | unnamed (Portugal) | • Proposing an optimization tool for the planning and operation of battery energy storage systems under | • The predominant factor in selecting the optimal BESS is the high investment costs, outweighing technical and technological | • Analysis assumes spinning reserve requirements without categorizing them by their activation |
| | | | | high RES penetration levels.<br>• Determining the optimal battery storage size, technology, and location. | features.<br>• Despite its relatively small size (4.3% peak load discharge capacity), introducing the appropriate BESS can significantly increase wind power penetration (35.7% curtailment reduction).<br>• As more non-flexible generators are added to the system, the reduction in fuel costs facilitated by the battery also increases. | timeframes.<br>• Analysis is focused on relatively low RES penetration levels. |
| [39] | 2016 | WFs, PVs | Pumped hydro | Crete (*Greece*) | • Examine the impact of increased RES penetration levels on the generation scheduling of the island system and further investigate whether pumped hydro storage station can mitigate the negative effects of enhanced renewables integration. | • PHS leads to a substantial decrease in annual production costs, particularly as RES penetration rises.<br>• The introduction of PHS does not result in significant changes in CO2 emissions. | • Study examines a single PHS configuration, without assessing its economic viability.<br>• Storage is almost solely used for energy arbitrage.<br>• The analytical mathematical formulation of the software is missing. |
| [47] | 2018 | WFs, PVs | Pumped hydro | Gran Canaria (*Spain*) | • Pathway towards full decarbonization of energy sectors in island systems, including electricity, heating, cooling, transportation.<br>• Investigation of the feasibility of achieving 100% RES penetration levels in all individual sectors and dimensioning of renewable capacity to achieve so. | • The potential to achieve a 75.9% RES penetration utilizing mature technologies and strategies that are currently deployable is proved.<br>• It is more advantageous to complement the introduction of PHS with an increase in installed PV capacity compared to increasing wind power capacity. | • Storage sizing assessment is limited to the power capacity.<br>• The employed EnergPLAN software is not adequately described. |
| [13] | 2018 | WFs, PVs | batteries hydrogen | Favignana (*Italy*) | • Identify the most economically favorable combination between battery and hydrogen storage facilities that reduces the obtained greenhouse gas emissions.<br>• Coupling between electricity and transportation sectors of the island<br>• Examine the sensitivity of the obtained results over the variations of several uncertain parameters to draw solid conclusions | • Combining batteries and an electrolyzer, is the most effective way to manage excess electricity compared to the development of a single storage technology, resulting in a higher carbon avoidance value and a lower LCOE.<br>• The system maintains its economic advantage even under adverse conditions, while the hybrid system fulfills the entire hydrogen-compressed natural gas public transport demand without requiring imports in all sensitivity cases.<br>• The renewable fraction and carbon avoidance metrics consistently yield positive results when compared to the island's current generation mix. | • RES penetration levels remain consistently low across all examined scenarios, (<14%)<br>• Hydrogen storage was exclusively utilized for transportation purposes.<br>• The study did not offer insights into the power capacity of BESS |
| [80] | 2018 | WFs, PVs | batteries | *not specified* | • Proposing a generation scheduling technique for incorporating storage in the UC-ED.<br>• Evaluating the feasibility of battery storage stations for implementation in island grids and proposing suitable remuneration schemes. | • BESS primarily provide primary up regulation reserves, while energy arbitrage is less favorable due to the quite uniform conventional generation mix in terms of its variable cost. This suggests that BESS configurations with limited storage capacity are the most viable options.<br>• In island systems with high conventional generation costs, BESS projects can be viable investments if their full potential is exploited through advanced system management tools and all associated benefits are recognized and taken into consideration. | • BESS provides reserves by constantly retaining a minimum energy charge equivalent to 30 minutes of operation.<br>• The stochasticity of RES generation is not accounted for. |
| [124] | 2018 | WFs, geothermal, waste | pumped hydro | *not specified* (European Atlantic Ocean) | • Investigating the potential of the pumped-hydro storage to provide frequency stability services to amplify the RES penetration to the island. To this end, the design of the pumps is investigated (fixed or variable speed) and numerous dynamic simula- | • The hydropower installations have limited impact on system frequency regulation when operating in turbining mode.<br>• In the event of severe grid disturbances, additional flexibility assets are necessary to offer frequency regulation services and stabilize system dynamics. | • Analysis does not consider sudden deviations in RES generation. |
| | | | | tions are performed. | • The challenges related to system stability caused by the PHS can be addressed with appropriate technical design modifications, without significantly complicating the system operator's actions. | |
| [81] | 2018 | WFs, PVs | batteries | *not specified* (2 island systems) | • Investigating the economic and operating system benefits deriving from the integration of battery storage into two island systems of different size. | • BESS facilitate the substitution of thermal energy with cost-effective wind production and enhance the efficiency of thermal units by operating at higher loads.<br>• In the case of a very small island, BESS play a crucial role in ensuring system security by providing the necessary primary up reserves to consistently meet system requirements. | • Only 1-h duration BESS configurations were investigated.<br>• An investment viability analysis for BESS is missing. |
| [185] | 2019 | WFs, PVs, biomass | batteries | Cyprus | • Examine the introduction of different battery technologies on system cost reduction and minimization of spinning reserves requirements.<br>• Economic evaluation of seven different battery energy storage types over their lifetime to conclude on the most favorable for system economics | • The choice of the optimal battery energy storage is predominantly influenced by the capital cost of the alternative choices.<br>• The zinc-air battery offers the highest net present value, whereas the lead-acid and sodium-sulfur battery systems are feasible investments. The Li-ion batteries are less preferred due to their high capital costs. | • BESS are the only candidates to fulfill spinning reserves requirements outside the UC-ED optimization.<br>• The optimization horizon for the analysis is restrained to one week. |
| [176] | 2019 | WFs, PVs | batteries flywheels | King, Flinders, Rottnest (*Tasmania/ Australia*) | • Presenting three actual island systems experiencing high-RES penetration levels under different approaches regarding the energy mix of each island. | • Highlighting the superiority of diesel low load operation is considerably more efficient for system economics against other solutions that include storage (batteries, flywheels). | • Model description lacks specific information regarding the operational constraints considered in the analysis.<br>• The increased wear-and-tear for diesel units operating under low-loading levels is not sufficiently evaluated. |
| [165] | 2020 | PVs | batteries, batteries integrated in electric boats | not specified (*island community*) | • Proposing a robust control strategy for battery energy storages and electrical boats (incorporating batteries and PVs) of isolated communities so that load-leveling and ancillaries, including frequency dynamics and fault ride through, to be provided to the grid. | • The proposed system demonstrates the ability to operate smoothly under various events when the island assets are effectively coordinated.<br>• Experimental testing in a laboratory-scale microgrid confirmed the robust behavior of the control algorithm. | • The experimental setup was designed for a single-phase system, and only sudden interruptions of sources and loads were tested, with three-phase fault conditions not included in the evaluation. |
| [100] | 2020 | WFs, PVs | batteries | Crete (*Greece*) | • Contribution of standalone battery energy storage to resource adequacy of an island system<br>• Comparison between a probabilistic and a deterministic methodology for evaluating capacity credit of batteries | • Regardless of the method examined, storage stations with limited rated power and increased energy capacity can achieve a capacity credit close to 100%.<br>• Large ESS configurations in terms of power capacity contribute to capacity adequacy with only a fraction of their rated power. | • Analysis assumes a single-dimensional operating pattern for storage to achieve load-leveling over the daily load curve. |
| [43] | 2020 | WFs, PVs, geotherm al | batteries | Faial (*Portugal*) | • Develop a multi-objective optimization model for an island microgrid towards high RES penetration levels, in a cost effective and reliable manner | • Introducing BESS provides the opportunity to increase RES penetration, with the increase proportionate to the RES capacity in the system.<br>• Demonstrating that the inclusion of carbon taxes elevates optimal storage and RES capacities. | • The study neglects the consideration of ramp constraints for thermal units.<br>• Reserve requirements are addressed through the implementation of spinning reserves. |
| [178] | 2020 | WFs, PVs | batteries | King Island (*Tasmania*) | • Fuel-efficient control strategies of all island assets are proposed and implemented within the tertiary control scheme. Three objectives are compared: | • Achieving annual RES penetration levels exceeding 60% in small island systems is possible with the right control strategy.<br>• Battery storage stations are less cost-competitive than adopting | • Study does not emphasize normal operation events.<br>• Dynamic models simulate primary and secondary |
| | | | | minimization of operating costs, maximization of renewable penetration levels, and achievement of zero-diesel operation via the low loading operation of diesel units. | low-loading operation for diesel units. | frequency control; however, the corresponding results are omitted. |
| [60] | 2020 | WFs, PVs | batteries flywheels | King Island, Molokai, (*Tasmania/ Hawaii*) | • Examine the potential of implementing diesel low-loading operation in two island systems of different characteristics to reduce the storage requirements for attaining high RES penetration levels and increase the RES hosting capacity. | • Low-load diesel operation can lead to a significant reduction in the system's battery requirements (43% and 50% calculated for the two examined systems) and a decrease in annual fuel consumption. | • Analysis oversimplifies the modeling of fuel curves for diesel generation, primarily when units operating below their technical minima. |
| [193] | 2020 | WFs, PVs | batteries | Galapagos (*Ecuador*) | • Determine the optimal configuration and generation schedule of island system over a project lifetime of 20-y and estimate the feasibility of wind-pv-batteries investment utilizing the LCOE metric. | • Highlighting the significant impact of meteorological conditions and economic parameters on the optimal solution. <br>• The optimal solution prioritizes higher capacities of PV and BESS combinations over WFs. <br>• BESS ensures flexibility and continuous service during adverse meteorological conditions. | • A rather simplistic approach is adopted for system assets' representation, ignoring various technical characteristics, such as unit ramp constraints. |
| [62] | 2020 | WFs, PVs | batteries | Baltra–Santa Cruz (*Ecuador*) | • Identify the pathway towards decarbonizing the island systems under investigation by examining several different scenarios that include the installation of additional RES capacity and the application of energy efficiency measures | • The development of distributed generation, like rooftop PV, leads to a reduction of over 50% in electricity bills. <br>• Due to the limited availability of land, distributed generation is viewed as the most attractive choice for zeroing fossil fuel consumption, in comparison to developing grid-scale RES or applying energy efficiency measures. | • Operating reserve remain constant over time at 15% of demand. <br>• BESS sizing ignores the power capacity of batteries. |
| [192] | 2020 | WFs | Pumped hydro | Terceira (*Portugal*) | • Develop an optimization model that determines the sizing of pumped hydro storage systems in terms of required pumping units and turbines to better exploit any RES surplus for the island system. <br>• The hydro-pumped storage station qualify as optimal are endogenously evaluated by the algorithm regarding their investment feasibility via the NPV and IRR metrics. Station revenues are the avoided variable system costs. | • Highlighting that the most influential factor affecting the Net Present Value of the projects is the availability of wind energy. <br>• It's also worth noting that the uncertainty surrounding CAPEX can have a significant impact. When CAPEX of PHS increases by 50% and is combined with a shortened project lifetime, it may result in a negative NPV. | • Constant efficiency values were employed for both turbines and pumps. However, using efficiency curves, especially for the pumps, would provide more accurate results, considering the substantial efficiency drop at lower power levels. <br>• The estimation of the infrastructure costs does not take into account a separate parcel for the penstock, depending on the length and materials. |
| [171] | 2021 | WFs, PVs | batteries | unnamed (small southern European island - 1.5 MW peak load) | • A load levelling algorithm for the operation of storage is proposed and the battery capacity is also determined, while accounting for the degradation of the storage asset. | • Demonstrating that the incorporation of BESS significantly enhances the performance of conventional units, reduce electricity costs, and facilitates the integration of additional RES. Specifically, BESS prevent thermal power plants from operating near their technical limits, thus minimize the potential for renewable energy curtailments. | • A single-dimensional operating strategy is assumed for BESS performing exclusively load levelling, without considering the bundle of services BESS can offer. |
| [166] | 2021 | WFs, PVs | batteries | unnamed (*European Atlantic Ocean*) | • Investigate the impact of load (mainly inductive motors) dynamics on storage sizing when the island operates with 100% inverter-based assets. | • Inductive motor loads have a negative impact on network dynamics due to their significant reactive power consumption following fault clearance. To address this issue, it is recommended to over-dimension the battery energy storage station. <br>• Revision of fault ride-through requirements in island power | • The study primarily focuses on a three-phase fault scenario and does not explore cases involving the unexpected loss of the largest online generator, sudden changes in load or RES generation. |
| | | | | | | systems may be necessary to accommodate inverter-based assets under inductive motor load conditions.<br>• The proposed fault ride-through control for converter-based RES could reduce the need for oversized storage to provide network regulation capabilities. | |
| [167] | 2021 | PVs | batteries | Cimei (*Republic of China-Taiwan*) | • Investigate control strategies of battery energy storage stations to deal with the intermittency of RES in island systems with high RES penetration levels. | • The suggested approach for handling sudden events minimizes frequency drops and prevents load shedding.<br>• Implementing the moving average control scheme for PV production smoothing significantly reduces PV ramping, lowering it from 15.1% to 0.29%.<br>• The proposed peak shaving strategy for BESS operation enhances the load factor of the microgrid system, improving it from 0.64 to 0.70. | • The study solely evaluates one BESS configuration without exploring alternative solutions.<br>• The primary emphasis is placed on addressing PV generation fluctuations, without investigating potential wind production interruptions. |
| [63] | 2021 | WFs, PVs | batteries | 147 Philippine off-grid islands (*Philippines*) | • Investigate the possibility of achieving high RES penetration rates in an economically viable manner the islands of Philippines and examine the pathways towards this transition. | • Hybrid grids incorporating solar and wind energy can potentially reduce electricity costs by 34.03% compared to diesel units dominated systems while achieving a 58.58% RES share in Philippine off-grid islands.<br>• The sensitivity analysis demonstrates that the reliance on solar and wind power in these islands remains robust in the face of uncertainties related to component costs and electricity demand. | • Study assumed that all installations are completed at the start of the project and would continuously supply the demand over the 20-year duration of the project.<br>• Only Li-ion BESS are assumed as possible choice for energy storage systems. |
| [101] | 2021 | WFs, PVs | batteries | *not specified* | • Estimating the contribution of battery energy systems to resource adequacy and computing their capacity credit by employing three operating strategies and Monte-Carlo simulations | • Employing a hybrid operation strategy, where the BESS perform daily peak shaving during high-demand days, while operating freely the rest of the time to minimize system operating cost, can strike a balanced compromise between system economics, capacity adequacy, and BESS lifetime expectancy compared to single-dimensional operation strategies. | • The daily peak shaving strategy for BESS is predetermined based on daily peak load forecasts and it is not dynamically decided internally to the Monte-Carlo model.<br>• The model does not consider the impact of real-time contingencies on BESS operation. |
| [99] | 2021 | WFs | batteries | *not specified* | • Compare the system benefit deriving from the introduction of HPS or centrally managed storage facilities in island systems and conclude on the most favorable storage scheme. | • Both storage management concepts support RES penetration levels over 45% annually.<br>• The centrally dispatched BESS design is more cost-efficient, requiring a smaller battery capacity than the HPS solution, leading to a reduced LCOE for achieving specific RES penetration targets and lowering the island system's overall generation cost. | • Regarding RES only the development of WFs is investigated.<br>• The analysis considers only BESS, with no exploration of other storage technologies. |
| [186] | 2022 | WFs, PVs | Batteries, pumped hydro | unnamed (*Tunisia*) | • Estimating the optimal system size in terms of number of batteries, wind turbines, PV panels and water reserved volume using a bi-objective optimization methodology. | • A Pareto front is established to choose balanced options between system cost and loss of power supply probability (LPSP), revealing substantial variations in the outcomes and storage sizing for each scenario. | • A rather simplified approach is adopted to represent the system and ESS operation.<br>• The criteria to determine the compromise zone and the chosen balance between LPSP and system cost could be presented with more clarity. |
| [179] | 2022 | WFs | Flywheels | San Cristobal (*Galapagos Islands,* | • Assessing the benefits of incorporating flywheel energy storage systems (FESS) in an NII microgrid in terms of frequency stability.<br>• Demonstrating the effectiveness of flywheels in | • Under normal operating conditions, the frequency variability due to an injection of intermittent wind generation is reduced by 46% with the proposed strategy for FESS and conventional generation, resulting in a 38% decrease in the average RoCoF | • The impact of various parameters, including RES penetration levels, mix, and flywheel sizing on FESS contribution to power system stability, have not been thoroughly explored. |
| | | | | *Ecuador)* | reducing frequency variability and improving the resilience of the microgrid in real-time, particularly in the presence of intermittent wind generation. | for the simulated time period.<br>• When the system is subjected to a contingency, the frequency nadir is reduced by 31%. | |
| [189] | 2022 | WFs, PVs | Batteries, pumped hydro | Grand Canary *(Spain)* | • Focusing on achieving a zero-emission generation system based entirely on renewable energies with 100% demand coverage and high system reliability for standalone grids.<br>• Analyzing three future scenarios, assessing the use of solar PV and wind technologies with different storage options to achieve an economically competitive and reliable generation mix. | • The demand load is met exclusively by RES generation (PV and wind) in a relative balanced way in the three scenarios,<br>• PHS is selected in all the scenarios, while BESS only in fully electrification scenario.<br>• The system is oversized, producing more energy than needed, leading to significant energy wastage in all three scenarios (approximately 30%). The installed power to peak consumption ratio ranges from 7 to 10. | • The capacity expansion model does not optimize the storage energy capacity independently from the storage power capacity. |
| [190] | 2022 | WFs, PVs | Batteries, hydrogen | Froan islands *(Norway)* | • Performing an enviromental analysis.<br>• Comparing the hydrogen-based system with alternative scenarios based on diesel generators and on the replacement of the outdated sea cable with a new one | • Interconnection with the Norwegian continental system results in slightly lower greenhouse gas emissions compared to a storage-based solution, but this conclusion may change with increased interconnection distance.<br>• The storage-based solution proves to be the most cost-effective.<br>• Relying solely on diesel generators appears to be the most expensive option and results in the highest GHG emissions.<br>• Stand-alone hydrogen-battery energy systems show great promise both economically and environmentally when compared to traditional configurations based on fossil fuels or grid connections. | • The study used the HyOpt tool for energy simulation and to determine the cost-optimal operation of the system, but the description lacks detailed information regarding the principles of the software. |
| [173] | 2022 | WFs, PVs | Batteries | São Vicente *(Cape Verde)* | • Applying a capacity expansion model for a long-term period up to 20 years for different scenarios of RES penetration levels (up to 100%) | • Aiming at 100% RES shares results in a substantial rise in system costs compared to the economically optimal scenario, due to the over-installation of RES, driven by their extreme seasonality in available generation and the high cost of energy storage.<br>• The optimal scenario stands out as the economically superior choice achieving a RES penetration level of approximately 90%. | • Only one representative week per month of each year is utilized to make planning decisions.<br>• Energy-to-power ratio of BESS is not optimized and remains predefined at 4-h BESS. |
| [191] | 2022 | WFs, PVs | Batteries, hydrogen | three islands *(Canada)* | • Designing of three hybrid energy systems to achieve the electrification of 50 residential buildings and the fueling of 25 hydrogen-powered cars.<br>• Providing a sensitivity analysis on project lifetime, capital cost multiplier for each component and salvage ratios to determine optimal solutions and major design sensitivities. | • Fuel cell costs have a higher impact on the LCOE compared to other components.<br>• The analysis of resource assessment volatility highlights the difficulty in predicting energy costs for short-term projects. | • Analysis does not detail the representation of the power system. |
| [158] | 2022 | WFs, PVs, biomass, CSP | Batteries, pumped hydro | Cyprus | • Investigating the storage needs for Cyprus to support a significant proportion of RES.<br>• Conducting a cost-benefit analysis to determine the optimal storage size and mixture. | • BESS configurations with 2-h or 3-h durations and power capacities at approximately 10% of peak load are the cost-optimal choices, yielding economic benefits of around 2.7 to 7 million euros, depending on the consideration of ESS contribution to adequacy. This leads to RES curtailments below 4% and RES penetration levels around 30-31%.<br>• The introduction of PHS becomes justified when its value in | • Only one configuration of PHS is examined.<br>• RES penetration levels up to ~32% are investigated. |
| | | | | | | terms of contributing to adequacy is considered. | |
| [180] | 2022 | WFs, PVs,geothermal, wave | Batteries | Pantelleria *(Italy)* | • Assessing the optimal RES mix under targeting at different RES penetration levels.<br>• Calculating the inherent response of the diesel generators and assessing contribution of BESS facilities to improve system stability. | • Higher RES penetration reduces the LCOE of the system but jeopardizes the system's rotational inertia, resulting in excessive frequency deviations during power imbalances.<br>• The study indicates that it is preferable to set a minimum synthetic inertia of 10 seconds for the BESS to keep the storage system size around 1 MW, as recommended by the optimal generation mix. | • All calculations are based on a representative day for each month throughout the year.<br>• An expansion model is employed with the goal of minimizing LCOE, considering solely the active power equilibrium.<br>• A modular approach is adopted, and the authors assumed specific RES rated power values for each technology. |
| [187] | 2022 | PVs | Batteries | South Andaman *(India)* | • Proposing a two-stage stohastic generation scheduling model including technical constraints that account for the interaction between PVs and BESS. | • PV primarily affects energy scheduling, while BESS is essential for reserve scheduling.<br>• Expanding PV from 5 MW to 50 MW has strong justification, but there's limited rationale for increasing BESS capacity from the initial 12 MWh to the newly proposed 25 MWh as higher BESS capacity doesn't lower overall energy and reserve scheduling costs or significantly reduce greenhouse gas emissions. | • The analysis only addresses spinning reserve requirements without classifying them into different types.<br>• Three options for BESS energy capacity are examined. |
| [1] | 2023 | WFs, PVs, Hydro, biomass | Batteries, pumped hydro | Gran Canary islands *(Spain)* | • Determining the optimal mix of generation and transmission capacity to satisfy energy demand at least cost.<br>• Considering the use of thermal and renewable generation, electric vehicles providing electricity-system services, BESS, pumped-hydro storage, HVAC and HVDC transmission lines between clusters of NII systems. | • Without considering the social cost of carbon, fossil-fueled generation is the predominantly used technology, while the establishment of carbon pricing increases RES share, up to a maximum of around 40%.<br>• Transmission lines development results in modest cost savings of 3% when compared to scenarios with no transmission investment.<br>• The utilization of energy storage leads to substantial reductions of up to 50% in required generation capacity and 23% in total costs.<br>• EVs primarily contribute to the electricity system by providing operating reserves.<br>• A combination of all these technologies yields cost savings of approximately 25% compared to current system costs. | • Generation investments are flexible, while transmission investments are fixed. A predefined set of candidate transmission lines with specified capacities is either constructed or not.<br>• Three types of operating reserves are accounted for: upward spinning, downward spinning, and upward non-spinning reserves.<br>• Electricity-system operation is represented using clustered weeks. |
| [21] | 2023 | WFs, PVs | Batteries, hydrogen | Pantelleria *(Italy)* | • Updating an existing capacity expansion model by applying a time series clustering method using representative days to enhance the modeling of energy systems that have a significant proportion of renewable energy sources.<br>• Various EES configurations, including batteries and hydrogen, were examined to reveal the specific features of each technology on renewable energy-supplied power systems. | • A hybrid development including both hydrogen storage and BESS is the most cost-effective option to achieve 100% RES penetration.<br>• The cost of a battery-only configuration is significantly higher, amounting to 155% more than the hydrogen-based scenarios.<br>• If only BESS are considered, the installed capacity of RES would need to be 5-6 times the peak load to attain 100% RES penetration. However, the introduction of hydrogen storage reduces this ratio to 2-2.5. | • Employs representative days for simulation to reduce the required simulation time while preserving accurate estimations of the objective function.<br>• BESS energy-to-power ratio is restricted by authors from 0.5 to 2 hours. |
| [188] | 2023 | WFs, PVs, tidal | Hydrogen | Larak island *(Iran)* | • Investigating the RES and storage needs of an island to reduce its dependancy on fossil fules.<br>• Examining different scenarios regarding the | • Integrating RES and storage can reduce total net current costs and diesel generator fuel consumption by 34% and 66%, respectively. | • Critical operating constraints, such as system reserve requirements and thermal units' operational limitations, are not considered. |

| | | | | | | |
|---|---|---|---|---|---|---|
| | | | | generation fleet of the island aiming at the minimization of a) the total system cost, b) the loss of power supply probability and c) the diesel generator fuel consumption. | • The authors suggest that choosing a very low, but non-zero, loss of power supply probability (e.g. 0.02) results in the highest reduction in total net present cost. | |
| [174] | 2023 | WFs, PVs | Batteries, pumped hydro | Santiago *(Cape Verde)* | • Expanding the work of [173] by applying the same methodology to a large island system that includes batteries, pumped-hydro storage and EVs as energy storage. Demand response techniques and sector coupling are also used to provide increased flexibility. | • Across the various scenarios, PV and PHS consistently held the highest installed power in the energy system, primarily influenced by economic considerations.<br>• While the optimal system development does not reach a 100% RES share, it is possible to achieve a fully decarbonized system by incorporating demand response capabilities, albeit at a slightly higher cost compared to the optimal scenario. | • Similar to [173], the analysis simulates only 12 indicative weeks.<br>• The proposed BESS configurations might reach an energy-to-power ratio of 21 hours. |

# Appendix C: Abbreviations

| | | |
|---|---|---|
| ASC | : | *Avoided system cost [€]* |
| BES | : | *Battery energy storage* |
| BESS | : | *Battery energy storage stations* |
| $C^{ess-charge}$ | : | *Cost of storage charging [€]* |
| capex | : | *Capital expenditure [€]* |
| CC | : | *Capacity credit* |
| CoS | : | *Cost of service [€]* |
| CP | : | *Capacity payment [€/MW]* |
| CR | : | *Capacity remuneration fee [€]* |
| $E^{ess}$ | : | *Annual energy storage discharge [MWh]* |
| EENS | : | *Expected energy not served* |
| ESS | : | *Energy storage stations* |
| FiT | : | *Feed-in-Tariff* |
| *i* | : | *Discount rate [%]* |
| HPS | : | *Hybrid power station* |
| HPS-O | : | *Hybrid power station operator* |
| LCOE | : | *Levelized cost of electricity* |
| LCOS | : | *Levelized cost of storage [€/MWh]* |
| LCOR | : | *Levelized cost of reserves [€/MWh]* |
| LOLE | : | *Loss of load expectation* |
| LP | : | *Linear programming* |
| LPSP | : | *Loss of power supply probability* |
| MILP | : | *Mixed integer linear programming* |
| NII | : | *Non-interconnected island* |
| opex | : | *Operational expenses [€]* |
| SO | : | *System Operator* |
| SoC | : | *State-of-charge* |
| $P^{ess}_{cc}$ | : | *Capacity credit of storage [MW]* |
| $R^{ess}$ | : | *Annual allocated reserves to storage [MWh]* |
| PHS | : | *Pumped hydro storage* |
| PV | : | *Photovoltaics* |
| RES | : | *Renewable energy sources* |
| UC-ED | : | *Unit commitment and economic dispatch* |
| $VC^{\text{w/o storage}}$ | : | *Variable system cost without storage [€]* |
| $VC^{\text{w/ storage}}$ | : | *Variable system cost with storage [€]* |
| VPP | : | *Virtual power plant* |
| WF | : | *Wind farms* |
| *y* | : | *Evaluation period [years]* |

# References

[1] J. Barrera-Santana, R. Sioshansi, An optimization framework for capacity planning of island electricity systems, Renew. Sustain. Energy Rev. 171 (2023) 112955. doi:10.1016/j.rser.2022.112955.

[2] L. Sigrist, E. Lobato, F.M. Echavarren, I. Egido, L. Rouco, Island Power Systems, CRC Press, Taylor & Francis Group, 6000 Broken Sound Parkway NW, Suite 300, Boca Raton, FL 33487-2742, 2016. doi:10.1201/9781315368740.

[3] E. Michalena, J.M. Hills, Paths of renewable energy development in small island developing states of the South Pacific, Renew. Sustain. Energy Rev. 82 (2018) 343–352. doi:10.1016/j.rser.2017.09.017.

[4] Y. Qiblawey, A. Alassi, M. Zain ul Abideen, S. Bañales, Techno-economic assessment of increasing the renewable energy supply in the Canary Islands: The case of Tenerife and Gran Canaria, Energy Policy. 162 (2022) 112791. doi:10.1016/j.enpol.2022.112791.

[5] G.N. Psarros, S.A. Papathanassiou, A unit commitment method for isolated power systems employing dual minimum loading levels to enhance flexibility, Electr. Power Syst. Res. 177 (2019) 106007. doi:10.1016/j.epsr.2019.106007.

[6] Y. Kuang, Y. Zhang, B. Zhou, C. Li, Y. Cao, L. Li, L. Zeng, A review of renewable energy utilization in islands, Renew. Sustain. Energy Rev. 59 (2016) 504–513. doi:10.1016/j.rser.2016.01.014.

[7] T.F. Agajie, A. Ali, A. Fopah-Lele, I. Amoussou, B. Khan, C.L.R. Velasco, E. Tanyi, A Comprehensive Review on Techno-Economic Analysis and Optimal Sizing of Hybrid Renewable Energy Sources with Energy Storage Systems, Energies. 16 (2023) 642. doi:10.3390/en16020642.

[8] A.I. López, A. Ramírez-Díaz, I. Castilla-Rodríguez, J. Gurriarán, J.A. Mendez-Perez, Wind farm energy surplus storage solution with second-life vehicle batteries in isolated grids, Energy Policy. 173 (2023) 113373. doi:10.1016/j.enpol.2022.113373.

[9] O. Erdinc, N.G. Paterakis, J.P.S. Catalão, Overview of insular power systems under increasing penetration of renewable energy sources: Opportunities and challenges, Renew. Sustain. Energy Rev. 52 (2015) 333–346. doi:10.1016/j.rser.2015.07.104.

[10] A.G.M.B. Mustayen, M.G. Rasul, X. Wang, M. Negnevitsky, J.M. Hamilton, Remote areas and islands power generation: A review on diesel engine performance and emission improvement techniques, Energy Convers. Manag. 260 (2022) 115614. doi:10.1016/j.enconman.2022.115614.

[11] G.N. Psarros, S.I. Nanou, S. V Papaefthymiou, S.A. Papathanassiou, Generation scheduling in non-interconnected islands with high RES penetration, Renew. Energy. 115 (2018) 338–352. doi:10.1016/j.renene.2017.08.050.

[12] G. Psarros, S. Papathanassiou, Comparative Assessment of Priority Listing and Mixed Integer Linear Programming Unit Commitment Methods for Non-Interconnected Island Systems, Energies. 12 (2019) 657. doi:10.3390/en12040657.

[13] D. Groppi, D. Astiaso Garcia, G. Lo Basso, F. Cumo, L. De Santoli, Analysing economic and environmental sustainability related to the use of battery and hydrogen energy storages for increasing the energy independence of small islands, Energy Convers. Manag. 177 (2018) 64–76. doi:10.1016/j.enconman.2018.09.063.

[14] S.-W. Hwangbo, B.-J. Kim, J.-H. Kim, Application of economic operation strategy on battery energy storage system at Jeju, in: 2013 IEEE PES Conf. Innov. Smart Grid Technol. (ISGT Lat. Am., IEEE, 2013: pp. 1–8. doi:10.1109/ISGT-LA.2013.6554367.

[15] K. Matsumoto, Y. Matsumura, Challenges and economic effects of introducing renewable energy in a remote island: A case study of Tsushima Island, Japan, Renew. Sustain. Energy Rev. 162 (2022) 112456. doi:10.1016/j.rser.2022.112456.

[16] G.N. Psarros, S.A. Papathanassiou, Generation scheduling in island systems with variable renewable energy sources: A literature review, Renew. Energy. 205 (2023) 1105–1124. doi:10.1016/j.renene.2023.01.099.

[17] J. Mitali, S. Dhinakaran, A.A. Mohamad, Energy storage systems: a review, Energy Storage Sav. 1 (2022) 166–216. doi:10.1016/j.enss.2022.07.002.

[18] M.Y. Worku, Recent Advances in Energy Storage Systems for Renewable Source Grid Integration: A Comprehensive Review, Sustain. 14 (2022). doi:10.3390/su14105985.

[19] Ayesha, M. Numan, M.F. Baig, M. Yousif, Reliability evaluation of energy storage systems combined with other grid flexibility options: A review, J. Energy Storage. 63 (2023) 107022. doi:10.1016/j.est.2023.107022.

[20] M. Eskandari, A. Rajabi, A. V. Savkin, M.H. Moradi, Z.Y. Dong, Battery energy storage systems (BESSs) and the economy-dynamics of microgrids: Review, analysis, and classification for standardization of BESSs applications, J. Energy Storage. 55 (2022). doi:10.1016/j.est.2022.105627.

[21] P. Marocco, R. Novo, A. Lanzini, G. Mattiazzo, M. Santarelli, Towards 100% renewable energy systems: The role of hydrogen and batteries, J. Energy Storage. 57 (2023) 106306. doi:10.1016/j.est.2022.106306.

[22] S. Koohi-Fayegh, M.A. Rosen, A review of energy storage types, applications and recent developments, J. Energy Storage. 27 (2020) 101047. doi:10.1016/j.est.2019.101047.

[23] Z. Topalović, R. Haas, A. Ajanović, M. Sayer, Prospects of electricity storage, Renew. Energy Environ. Sustain. 8 (2023) 2. doi:10.1051/rees/2022016.

[24] M.M. Rahman, A.O. Oni, E. Gemechu, A. Kumar, Assessment of energy storage technologies: A review, Energy Convers. Manag. 223 (2020) 113295. doi:10.1016/j.enconman.2020.113295.

[25] G.N. Psarros, S.A. Papathanassiou, Electricity storage requirements to support the transition towards high renewable penetration levels – Application to the Greek power system, J. Energy Storage. 55 (2022). doi:10.1016/j.est.2022.105748.

[26] D. Groppi, A. Pfeifer, D.A. Garcia, G. Krajačić, N. Duić, A review on energy storage and demand side management solutions in smart energy islands, Renew. Sustain. Energy Rev. 135 (2021). doi:10.1016/j.rser.2020.110183.

[27] G. Notton, Importance of islands in renewable energy production and storage: The situation of the French islands, Renew. Sustain. Energy Rev. 47 (2015) 260–269. doi:10.1016/j.rser.2015.03.053.

[28] E.M.G. Rodrigues, G.J. Osório, R. Godina, A.W. Bizuayehu, J.M. Lujano-Rojas, J.P.S. Catalão, Grid code reinforcements for deeper renewable generation in insular energy systems, Renew. Sustain. Energy Rev. 53 (2016) 163–177. doi:10.1016/j.rser.2015.08.047.

[29] C. Bueno, J.A. Carta, Wind powered pumped hydro storage systems, a means of increasing the penetration of renewable energy in the Canary Islands, Renew. Sustain. Energy Rev. 10 (2006) 312–340. doi:10.1016/j.rser.2004.09.005.

[30] J.K. Kaldellis, D. Zafirakis, K. Kavadias, Techno-economic comparison of energy storage systems for island autonomous electrical networks, Renew. Sustain. Energy Rev. 13 (2009) 378–392. doi:10.1016/j.rser.2007.11.002.

[31] F. Chen, N. Duic, L. Manuel Alves, M. da Graça Carvalho, Renewislands—Renewable energy solutions for islands, Renew. Sustain. Energy Rev. 11 (2007) 1888–1902. doi:10.1016/j.rser.2005.12.009.

[32] G. Caralis, K. Rados, A. Zervos, On the market of wind with hydro-pumped storage systems in autonomous Greek islands, Renew. Sustain. Energy Rev. 14 (2010) 2221–2226. doi:10.1016/j.rser.2010.02.008.

[33] G. Frydrychowicz-Jastrzębska, El Hierro Renewable Energy Hybrid System: A Tough Compromise, Energies. 11 (2018) 2812. doi:10.3390/en11102812.

[34] A. Ioannidis, K.J. Chalvatzis, X. Li, G. Notton, P. Stephanides, The case for islands' energy vulnerability: Electricity supply diversity in 44 global islands, Renew. Energy. 143 (2019) 440–452. doi:10.1016/j.renene.2019.04.155.

[35] G. Tsilingiridis, C. Sidiropoulos, A. Pentaliotis, Reduction of air pollutant emissions using renewable energy sources for power generation in Cyprus, Renew. Energy. 36 (2011) 3292–3296.

doi:10.1016/j.renene.2011.04.030.

[36] I. Miranda, N. Silva, H. Leite, A Holistic Approach to the Integration of Battery Energy Storage Systems in Island Electric Grids with High Wind Penetration, IEEE Trans. Sustain. Energy. 7 (2016) 775–785. doi:10.1109/TSTE.2015.2497003.

[37] K.J. Chua, W.M. Yang, S.S. Er, C.A. Ho, Sustainable energy systems for a remote island community, Appl. Energy. 113 (2014) 1752–1763. doi:10.1016/j.apenergy.2013.09.030.

[38] V. Cosentino, S. Favuzza, G. Graditi, M.G. Ippolito, F. Massaro, E. Riva Sanseverino, G. Zizzo, Smart renewable generation for an islanded system. Technical and economic issues of future scenarios, Energy. 39 (2012) 196–204. doi:10.1016/j.energy.2012.01.030.

[39] C.K. Simoglou, E.A. Bakirtzis, P.N. Biskas, A.G. Bakirtzis, Optimal operation of insular electricity grids under high RES penetration, Renew. Energy. 86 (2016) 1308–1316. doi:10.1016/j.renene.2015.09.064.

[40] E. Riva Sanseverino, R. Riva Sanseverino, S. Favuzza, V. Vaccaro, Near zero energy islands in the Mediterranean: Supporting policies and local obstacles, Energy Policy. 66 (2014) 592–602. doi:10.1016/j.enpol.2013.11.007.

[41] S. V. Papaefthymiou, E.G. Karamanou, S.A. Papathanassiou, M.P. Papadopoulos, A wind-hydro-pumped storage station leading to high RES penetration in the autonomous island system of Ikaria, IEEE Trans. Sustain. Energy. 1 (2010) 163–172. doi:10.1109/TSTE.2010.2059053.

[42] C. Bueno, J.A. Carta, Technical–economic analysis of wind-powered pumped hydrostorage systems. Part I: model development, Sol. Energy. 78 (2005) 382–395. doi:10.1016/j.solener.2004.08.006.

[43] M. Barbaro, R. Castro, Design optimisation for a hybrid renewable microgrid: Application to the case of Faial island, Azores archipelago, Renew. Energy. 151 (2020) 434–445. doi:10.1016/j.renene.2019.11.034.

[44] H.M. Trondheim, B.A. Niclasen, T. Nielsen, F.F. Da Silva, C.L. Bak, 100% Sustainable Electricity in the Faroe Islands: Expansion Planning through Economic Optimization, IEEE Open Access J. Power Energy. 8 (2021) 23–34. doi:10.1109/OAJPE.2021.3051917.

[45] F.A. Canales, J.K. Jurasz, M. Guezgouz, A. Beluco, Cost-reliability analysis of hybrid pumped-battery storage for solar and wind energy integration in an island community, Sustain. Energy Technol. Assessments. 44 (2021) 101062. doi:10.1016/j.seta.2021.101062.

[46] G. Barone, A. Buonomano, C. Forzano, G.F. Giuzio, A. Palombo, Increasing renewable energy penetration and energy independence of island communities: A novel dynamic simulation approach for energy, economic, and environmental analysis, and optimization, J. Clean. Prod. 311 (2021) 127558. doi:10.1016/j.jclepro.2021.127558.

[47] P. Cabrera, H. Lund, J.A. Carta, Smart renewable energy penetration strategies on islands: The case of Gran Canaria, Energy. 162 (2018) 421–443. doi:10.1016/j.energy.2018.08.020.

[48] C. Bueno, J.A. Carta, Technical-economic analysis of wind-powered pumped hydrostorage systems. Part II: Model application to the island of El Hierro, Sol. Energy. 78 (2005) 396–405. doi:10.1016/j.solener.2004.08.007.

[49] I. Kougias, S. Szabó, A. Nikitas, N. Theodossiou, Sustainable energy modelling of non-interconnected Mediterranean islands, Renew. Energy. 133 (2019) 930–940. doi:10.1016/j.renene.2018.10.090.

[50] E.I. Vrettos, S.A. Papathanassiou, Operating Policy and Optimal Sizing of a High Penetration RES-BESS System for Small Isolated Grids, IEEE Trans. Energy Convers. 26 (2011) 744–756. doi:10.1109/TEC.2011.2129571.

[51] D.M. Gioutsos, K. Blok, L. van Velzen, S. Moorman, Cost-optimal electricity systems with increasing renewable energy penetration for islands across the globe, Appl. Energy. 226 (2018) 437–449. doi:10.1016/j.apenergy.2018.05.108.

[52] D. Thomas, O. Deblecker, C.S. Ioakimidis, Optimal design and techno-economic analysis of an autonomous small isolated microgrid aiming at high RES penetration, Energy. 116 (2016) 364–379. doi:10.1016/j.energy.2016.09.119.

[53] M. Drouineau, E. Assoumou, V. Mazauric, N. Maïzi, Increasing shares of intermittent sources in Reunion Island: Impacts on the future reliability of power supply, Renew. Sustain. Energy Rev. 46 (2015) 120–128. doi:10.1016/j.rser.2015.02.024.

[54] N. Maïzi, V. Mazauric, E. Assoumou, S. Bouckaert, V. Krakowski, X. Li, P. Wang, Maximizing intermittency in 100% renewable and reliable power systems: A holistic approach applied to Reunion Island in 2030, Appl. Energy. 227 (2018) 332–341. doi:10.1016/j.apenergy.2017.08.058.

[55] N. Duić, M. Da Graça Carvalho, Increasing renewable energy sources in island energy supply: Case study Porto Santo, Renew. Sustain. Energy Rev. 8 (2004) 383–399. doi:10.1016/j.rser.2003.11.004.

[56] J. Kersey, P. Blechinger, R. Shirley, A panel data analysis of policy effectiveness for renewable energy expansion on Caribbean islands, Energy Policy. 155 (2021) 112340. doi:10.1016/j.enpol.2021.112340.

[57] P. Arévalo, A.A. Eras-Almeida, A. Cano, F. Jurado, M.A. Egido-Aguilera, Planning of electrical energy for the Galapagos Islands using different renewable energy technologies, Electr. Power Syst. Res. 203 (2022). doi:10.1016/j.epsr.2021.107660.

[58] H.M. Marczinkowski, P.A. Østergaard, S.R. Djørup, Transitioning island energy systems—Local conditions, development phases, and renewable energy integration, Energies. 12 (2019). doi:10.3390/en12183484.

[59] C.D. Yue, C.S. Chen, Y.C. Lee, Integration of optimal combinations of renewable energy sources into the energy supply of Wang-An Island, Renew. Energy. 86 (2016) 930–942. doi:10.1016/j.renene.2015.08.073.

[60] J. Hamilton, M. Negnevitsky, X. Wang, E. Semshchikov, The role of low-load diesel in improved renewable hosting capacity within isolated power systems, Energies. 13 (2020). doi:10.3390/en13164053.

[61] K. Fiorentzis, Y. Katsigiannis, E. Karapidakis, Full-scale implementation of res and storage in an island energy system, Inventions. 5 (2020) 1–14. doi:10.3390/inventions5040052.

[62] A.A. Eras-Almeida, M.A. Egido-Aguilera, P. Blechinger, S. Berendes, E. Caamaño, E. García-Alcalde, Decarbonizing the Galapagos Islands: Techno-economic perspectives for the hybrid renewable mini-grid Baltra-Santa Cruz, Sustain. 12 (2020). doi:10.3390/su12062282.

[63] J.D.A. Pascasio, E.A. Esparcia, M.T. Castro, J.D. Ocon, Comparative assessment of solar photovoltaic-wind hybrid energy systems: A case for Philippine off-grid islands, Renew. Energy. 179 (2021) 1589–1607. doi:10.1016/j.renene.2021.07.093.

[64] R.J. Brecha, K. Schoenenberger, M. Ashtine, R.K. Koon, Ocean thermal energy conversion—flexible enabling technology for variable renewable energy integration in the caribbean, Energies. 14 (2021). doi:10.3390/en14082192.

[65] R. Carapellucci, L. Giordano, Modeling and optimization of an energy generation island based on renewable technologies and hydrogen storage systems, Int. J. Hydrogen Energy. 37 (2012) 2081–2093. doi:10.1016/j.ijhydene.2011.10.073.

[66] M.H. Ashourian, S.M. Cherati, A.A. Mohd Zin, N. Niknam, A.S. Mokhtar, M. Anwari, Optimal green energy management for island resorts in Malaysia, Renew. Energy. 51 (2013) 36–45. doi:10.1016/j.renene.2012.08.056.

[67] M. Ghofrani, A. Arabali, M. Etezadi-Amoli, M.S. Fadali, A framework for optimal placement of energy storage units within a power system with high wind penetration, IEEE Trans. Sustain. Energy. 4 (2013) 434–442. doi:10.1109/TSTE.2012.2227343.

[68] P.D. Brown, J.A. Pecas Lopes, M.A. Matos, Optimization of Pumped Storage Capacity in an Isolated Power System With Large Renewable Penetration, IEEE Trans. Power Syst. 23 (2008) 523–531. doi:10.1109/TPWRS.2008.919419.

[69] S. Gill, G.W. Ault, I. Kockar, The optimal operation of energy storage in a wind power curtailment scheme, in: 2012 IEEE Power Energy Soc. Gen. Meet., IEEE, 2012: pp. 1–8. doi:10.1109/PESGM.2012.6344705.

[70] B. Kroposki, R. Lasseter, T. Ise, S. Morozumi, S. Papathanassiou, N. Hatziargyriou, Making microgrids work, IEEE Power Energy Mag. 6 (2008) 40–53. doi:10.1109/MPE.2008.918718.

[71] G. Delille, B. François, G. Malarange, Dynamic Frequency Control Support by Energy Storage to Reduce the Impact of Wind and Solar Generation on Isolated Power System's Inertia, IEEE Trans. Sustain. Energy. 3 (2012) 931–939. doi:10.1109/TSTE.2012.2205025.

[72] R. Sebastián, Application of a battery energy storage for frequency regulation and peak shaving in a wind diesel power system, IET Gener. Transm. Distrib. 10 (2016) 764–770. doi:10.1049/iet-gtd.2015.0435.

[73] Y. Wen, W. Li, G. Huang, X. Liu, Frequency Dynamics Constrained Unit Commitment with Battery Energy Storage, IEEE Trans. Power Syst. 31 (2016) 5115–5125. doi:10.1109/TPWRS.2016.2521882.

[74] K.C. Divya, J. Østergaard, Battery energy storage technology for power systems—An overview, Electr. Power Syst. Res. 79 (2009) 511–520. doi:10.1016/j.epsr.2008.09.017.

[75] D. McConnell, T. Forcey, M. Sandiford, Estimating the value of electricity storage in an energy-only wholesale market, Appl. Energy. 159 (2015) 422–432. doi:10.1016/j.apenergy.2015.09.006.

[76] W. He, L. Tao, L. Han, Y. Sun, E. Pietro Campana, J. Yan, Optimal analysis of a hybrid renewable power system for a remote island, Renew. Energy. 179 (2021) 96–104. doi:10.1016/j.renene.2021.07.034.

[77] G. Martínez-Lucas, J. Sarasúa, J. Sánchez-Fernández, Frequency Regulation of a Hybrid Wind–Hydro Power Plant in an Isolated Power System, Energies. 11 (2018) 239. doi:10.3390/en11010239.

[78] L. Sigrist, E. Lobato, L. Rouco, Energy storage systems providing primary reserve and peak shaving in small isolated power systems: An economic assessment, Int. J. Electr. Power Energy Syst. 53 (2013) 675–683. doi:10.1016/j.ijepes.2013.05.046.

[79] I. Egido, L. Sigrist, E. Lobato, L. Rouco, A. Barrado, P. Fontela, J. Magriñá, Energy storage systems for frequency stability enhancement in small-isolated power systems, in: Int. Conf. Renew. Energies Power Qual., 2015: pp. 1–6.

[80] G.N. Psarros, E.G. Karamanou, S.A. Papathanassiou, Feasibility Analysis of Centralized Storage Facilities in Isolated Grids, IEEE Trans. Sustain. Energy. 9 (2018) 1822–1832. doi:10.1109/TSTE.2018.2816588.

[81] G.N. Psarros, S.P. Kokkolios, S.A. Papathanassiou, Centrally Managed Storage Facilities in Small Non-Interconnected Island Systems, in: 2018 53rd Int. Univ. Power Eng. Conf., IEEE, 2018: pp. 1–6. doi:10.1109/UPEC.2018.8542102.

[82] G.N. Psarros, S.A. Papathanassiou, Evaluation of battery-renewable hybrid stations in small-isolated systems, IET Renew. Power Gener. 14 (2020) 39–51. doi:10.1049/iet-rpg.2019.0212.

[83] G.N. Psarros, S.A. Papathanassiou, Operation of a wind-PV-battery hybrid power station in an isolated island grid, in: Mediterr. Conf. Power Gener. Transm. Distrib. Energy Convers. (MEDPOWER 2018), Institution of Engineering and Technology, Dubrovnik, Croatia, 2018: pp. 4 (6 pp.)-4 (6 pp.). doi:10.1049/cp.2018.1836.

[84] H. Chen, T.N. Cong, W. Yang, C. Tan, Y. Li, Y. Ding, Progress in electrical energy storage system: A critical review, Prog. Nat. Sci. 19 (2009) 291–312. doi:10.1016/j.pnsc.2008.07.014.

[85] IEA, Bioenergy for Electricity and Heat, 2011.

[86] IRENA, IEA-ETSAP, Technology Brief - Electricity Storage, (2012) 28.

[87] ITRE, Outlook for Energy Storage Technologies, 2008.

[88] J. Leadbetter, L.G. Swan, Selection of battery technology to support grid-integrated renewable electricity, J. Power Sources. 216 (2012) 376–386. doi:10.1016/j.jpowsour.2012.05.081.

[89] P. Kundur, J. Paserba, V. Ajjarapu, G. Andersson, A. Bose, T. Van Cutsem, C. Canizares, N. Hatziargyriou, D. Hill, V. Vittal, A. Stankovic, C. Taylor, Definition and Classification of Power System Stability IEEE/CIGRE Joint Task Force on Stability Terms and Definitions, IEEE Trans. Power Syst. 19 (2004) 1387–1401. doi:10.1109/TPWRS.2004.825981.

[90] Chang, Lu, Wu, Frequency-regulating reserve constrained unit commitment for an isolated power system, IEEE Trans. Power Syst. 28 (2013) 578–586. doi:10.1109/TPWRS.2012.2208126.

[91] L.E. Sokoler, P. Vinter, R. Baerentsen, K. Edlund, J.B. Jorgensen, Contingency-Constrained Unit Commitment in Meshed Isolated Power Systems, IEEE Trans. Power Syst. 31 (2016) 3516–3526. doi:10.1109/TPWRS.2015.2485781.

[92] M. Carrión, R. Zárate-Miñano, F. Milano, Impact of off-nominal frequency values on the generation scheduling of small-size power systems, Int. J. Electr. Power Energy Syst. 122 (2020) 106174. doi:10.1016/j.ijepes.2020.106174.

[93] D.T. Lagos, N.D. Hatziargyriou, Data-Driven Frequency Dynamic Unit Commitment for Island Systems with High RES Penetration, IEEE Trans. Power Syst. 36 (2021) 4699–4711. doi:10.1109/TPWRS.2021.3060891.

[94] G. Graditi, M.G. Ippolito, E. Telaretti, G. Zizzo, An Innovative Conversion Device to the Grid Interface of Combined RES-Based Generators and Electric Storage Systems, IEEE Trans. Ind. Electron. 62 (2015) 2540–2550. doi:10.1109/TIE.2014.2336620.

[95] E. Liegmann, R. Majumder, An Efficient Method of Multiple Storage Control in Microgrids, IEEE Trans. Power Syst. 30 (2015) 3437–3444. doi:10.1109/TPWRS.2014.2385156.

[96] R. Sebastián, A. Nevado, Study and simulation of awind hydro isolated microgrid, Energies. 13 (2020) 1–15. doi:10.3390/en13225937.

[97] P.B.L. Neto, O.R. Saavedra, L.A. De Souza Ribeiro, A Dual-Battery Storage Bank Configuration for Isolated Microgrids Based on Renewable Sources, IEEE Trans. Sustain. Energy. 9 (2018) 1618–1626. doi:10.1109/TSTE.2018.2800689.

[98] G. Caralis, A. Zervos, Value of wind energy on the reliability of autonomous power systems, IET Renew. Power Gener. 4 (2010) 186. doi:10.1049/iet-rpg.2009.0052.

[99] G.N. Psarros, P.A. Dratsas, S.A. Papathanassiou, A comparison between central- and self-dispatch storage management principles in island systems, Appl. Energy. 298 (2021) 117181. doi:10.1016/j.apenergy.2021.117181.

[100] G.N. Psarros, A.G. Papakonstantinou, J.S. Anagnostopoulos, S.A. Papathanassiou, N.G. Boulaxis, Contribution of energy storage to capacity adequacy - Application to island power systems, in: CIGRE 48th Sess., Paris, France, 2020: pp. 1–10.

[101] P.A. Dratsas, G.N. Psarros, S.A. Papathanassiou, Battery Energy Storage Contribution to System Adequacy, Energies. 14 (2021) 5146. doi:10.3390/en14165146.

[102] M. Farrokhabadi, S. Konig, C.A. Canizares, K. Bhattacharya, T. Leibfried, Battery Energy Storage System Models for Microgrid Stability Analysis and Dynamic Simulation, IEEE Trans. Power Syst. 33 (2018) 2301–2312. doi:10.1109/TPWRS.2017.2740163.

[103] D.S. Miranda, Y. Sun, J.F.G. Cobben, M. Gibescu, Impact of energy storage on island grid dynamics: A case study of Bonaire, in: 2016 IEEE Int. Energy Conf., IEEE, 2016: pp. 1–7. doi:10.1109/ENERGYCON.2016.7513940.

[104] J. Gouveia, C.L. Moreira, J.A.P. Lopes, Rule-based adaptive control strategy for grid-forming inverters in islanded power systems for improving frequency stability, Electr. Power Syst. Res. 197 (2021) 107339. doi:10.1016/j.epsr.2021.107339.

[105] J. Gouveia, C.L. Moreira, J.A.P. Lopes, Grid-Forming Inverters Sizing in Islanded Power Systems- A stability perspective, SEST 2019 - 2nd Int. Conf. Smart Energy Syst. Technol. (2019) 19–24. doi:10.1109/SEST.2019.8849110.

[106] D.R. Fragoso, A. dos Santos, E.C. Fernandes, Increasing RES penetration through H2 technologies on Flores island, Azores: A techno-economic analysis, Int. J. Hydrogen Energy. (2023). doi:10.1016/j.ijhydene.2023.02.055.

[107] P.A. Dratsas, G.N. Psarros, S.A. Papathanassiou, Feasibility of Behind-the-Meter Battery Storage in Wind Farms Operating on Small Islands, Batteries. 8 (2022) 275. doi:10.3390/batteries8120275.

[108] L. Tziovani, L. Hadjidemetriou, S. Timotheou, Energy Scheduling of Wind-Storage Systems Using Stochastic and Robust Optimization, in: 2022 IEEE Power Energy Soc. Gen. Meet., IEEE, 2022: pp. 1–5. doi:10.1109/PESGM48719.2022.9916673.

[109] M.J. Aziz, D.F. Gayme, K. Johnson, J. Knox-Hayes, P. Li, E. Loth, L.Y. Pao, D.R. Sadoway, J. Smith, S. Smith, A co-design framework for wind energy integrated with storage, Joule. 6 (2022) 1995–2015. doi:10.1016/j.joule.2022.08.014.

[110] W. Wang, B. Yuan, Q. Sun, R. Wennersten, Application of energy storage in integrated energy systems — A solution to fluctuation and uncertainty of renewable energy, J. Energy Storage. 52 (2022) 104812. doi:10.1016/j.est.2022.104812.

[111] Z. Wang, G. Fang, X. Wen, Q. Tan, P. Zhang, Z. Liu, Coordinated operation of conventional hydropower plants as hybrid pumped storage hydropower with wind and photovoltaic plants, Energy Convers. Manag. 277 (2023) 116654. doi:10.1016/j.enconman.2022.116654.

[112] D. Pudjianto, C. Ramsay, G. Strbac, Virtual power plant and system integration of distributed energy resources, IET Renew. Power Gener. 1 (2007) 10. doi:10.1049/iet-rpg:20060023.

[113] L. Baringo, M. Rahimiyan, Virtual Power Plants and Electricity Markets, Springer International Publishing, Cham, 2020. doi:10.1007/978-3-030-47602-1.

[114] W. Nafkha-Tayari, S. Ben Elghali, E. Heydarian-Forushani, M. Benbouzid, Virtual Power Plants Optimization Issue: A Comprehensive Review on Methods, Solutions, and Prospects, Energies. 15 (2022) 3607. doi:10.3390/en15103607.

[115] H. Pandžić, I. Kuzle, T. Capuder, Virtual power plant mid-term dispatch optimization, Appl. Energy. 101 (2013) 134–141. doi:10.1016/j.apenergy.2012.05.039.

[116] C. Kieny, B. Berseneff, N. Hadjsaid, Y. Besanger, J. Maire, On the concept and the interest of virtual power plant: Some results from the European project Fenix, in: 2009 IEEE Power Energy Soc. Gen. Meet., IEEE, 2009: pp. 1–6. doi:10.1109/PES.2009.5275526.

[117] E.A. Bhuiyan, M.Z. Hossain, S.M. Muyeen, S.R. Fahim, S.K. Sarker, S.K. Das, Towards next generation virtual power plant: Technology review and frameworks, Renew. Sustain. Energy Rev. 150 (2021) 111358. doi:10.1016/j.rser.2021.111358.

[118] A.G. Zamani, A. Zakariazadeh, S. Jadid, Day-ahead resource scheduling of a renewable energy based virtual power plant, Appl. Energy. 169 (2016) 324–340. doi:10.1016/j.apenergy.2016.02.011.

[119] G. Caralis, A. Zervos, Analysis of the combined use of wind and pumped storage systems in autonomous Greek islands, IET Renew. Power Gener. 1 (2007) 49. doi:10.1049/iet-rpg:20060010.

[120] S. Papaefthimiou, E. Karamanou, S. Papathanassiou, M. Papadopoulos, Operating policies for wind-pumped storage hybrid power stations in island grids, IET Renew. Power Gener. 3 (2009) 293. doi:10.1049/iet-rpg.2008.0071.

[121] J.S. Anagnostopoulos, D.E. Papantonis, Simulation and size optimization of a pumped–storage power plant for the recovery of wind-farms rejected energy, Renew. Energy. 33 (2008) 1685–1694. doi:10.1016/j.renene.2007.08.001.

[122] A. V. Ntomaris, A.G. Bakirtzis, Stochastic Scheduling of Hybrid Power Stations in Insular Power Systems With High Wind Penetration, IEEE Trans. Power Syst. 31 (2016) 3424–3436. doi:10.1109/TPWRS.2015.2499039.

[123] D. Al Katsaprakakis, D.G. Christakis, K. Pavlopoylos, S. Stamataki, I. Dimitrelou, I. Stefanakis, P. Spanos, Introduction of a wind powered pumped storage system in the isolated insular power system of Karpathos–Kasos, Appl. Energy. 97 (2012) 38–48. doi:10.1016/j.apenergy.2011.11.069.

[124] P. Beires, M.H. Vasconcelos, C.L. Moreira, J.A. Peças Lopes, Stability of autonomous power systems with reversible hydro power plants, Electr. Power Syst. Res. 158 (2018) 1–14. doi:10.1016/j.epsr.2017.12.028.

[125] S. V Papaefthymiou, V.G. Lakiotis, I.D. Margaris, S.A. Papathanassiou, Dynamic analysis of island systems with wind-pumped-storage hybrid power stations, Renew. Energy. 74 (2015) 544–554.

doi:10.1016/j.renene.2014.08.062.

[126] S. V. Papaefthymiou, S.A. Papathanassiou, Optimum sizing of wind-pumped-storage hybrid power stations in island systems, Renew. Energy. 64 (2014) 187–196. doi:10.1016/j.renene.2013.10.047.

[127] Lazard's Levelized Cost of Storage Analysis - Version 4.0, (2018). https://www.lazard.com/media/450774/lazards-levelized-cost-of-storage-version-40-vfinal.pdf.

[128] D.L. Wood, J. Li, C. Daniel, Prospects for reducing the processing cost of lithium ion batteries, J. Power Sources. 275 (2015) 234–242. doi:10.1016/j.jpowsour.2014.11.019.

[129] Brinsmead T.S., Graham P., Hayward J., Ratnam E.L., Reedman L., Future Energy Storage Trends: An Assessment of the Economic Viability, Potential Uptake and Impacts of Electrical Energy Storage on the NEM 2015–2035, CSIRO Australia, 2015. http://www.aemc.gov.au/Major-Pages/Integration-of-storage/Documents/CSIRIO-Future-Trends-Report-2015.aspx.

[130] X. Hu, C. Zou, C. Zhang, Y. Li, Technological Developments in Batteries: A Survey of Principal Roles, Types, and Management Needs, IEEE Power Energy Mag. 15 (2017) 20–31. doi:10.1109/MPE.2017.2708812.

[131] TILOS Project – Eunice Energy Group, (n.d.). http://eunice-group.com/projects/tilos-project/.

[132] D. Al Katsaprakakis, Hybrid power plants in non-interconnected insular systems, Appl. Energy. 164 (2016) 268–283. doi:10.1016/j.apenergy.2015.11.085.

[133] A. V. Ntomaris, A.G. Bakirtzis, Optimal bidding for risk-averse hybrid power station producers in insular power systems: An MPEC approach, in: 2017 IEEE PES Innov. Smart Grid Technol. Conf. Eur., IEEE, Torino, Italy, 2017: pp. 1–6. doi:10.1109/ISGTEurope.2017.8260169.

[134] A. V. Ntomaris, A.G. Bakirtzis, Optimal Bidding of Hybrid Power Stations in Insular Power Systems, IEEE Trans. Power Syst. 32 (2017) 3782–3793. doi:10.1109/TPWRS.2016.2632971.

[135] G.N. Psarros, S.A. Papathanassiou, Internal dispatch for RES-storage hybrid power stations in isolated grids, Renew. Energy. 147 (2020) 2141–2150. doi:10.1016/j.renene.2019.10.001.

[136] Law 3468/2006, Generation of Electricity using Renewable Energy Sources and High-Efficiency Cogeneration of Electricity and Heat and Miscellaneous Provisions, Off. Gazzette A', p. 129/27-6-2006, 2006.

[137] Law 3851/2010, Accelerating the development of Renewable Energy Sources to deal with Climate Change and other regulations in topics under the authority of Ministry of Environment, Energy and Climate Change, Off. Gazzette A', p. 85/4-6-2010, 2010. http://www.lagie.gr/fileadmin/user_upload/Files/adeiodotisi/2010.06.04_FEK.3851_N.3851.pdf.

[138] European Commission, State aid SA.58482 - Remuneration scheme of Hybrid Power Stations in NIIs of Greece until 2026, Brussels, 21.12.2021, 2021.

[139] Hybrid Power Generation Systems | PPC, (n.d.). https://www.dei.gr/en/ppc-group/ppc/business-areas/renewable-energy-sources/hybrid-power-generation-systems/ (accessed April 26, 2023).

[140] E.E. Pompodakis, G.C. Kryonidis, E.S. Karapidakis, Volt/Var control and energy management in Non-Interconnected insular networks with multiple hybrid power plants, Appl. Energy. 331 (2023). doi:10.1016/j.apenergy.2022.120427.

[141] S. Karellas, N. Tzouganatos, Comparison of the performance of compressed-air and hydrogen energy storage systems: Karpathos island case study, Renew. Sustain. Energy Rev. 29 (2014) 865–882. doi:10.1016/j.rser.2013.07.019.

[142] N. Sakellaridis, J. Mantzaris, G. Tsourakis, C. Vournas, I. Vitellas, Operation and security assessment of the power system of Crete with integration of pumped storage and concentrated solar thermal plants, in: 2013 IREP Symp. Bulk Power Syst. Dyn. Control - IX Optim. Secur. Control Emerg. Power Grid, IEEE, 2013: pp. 1–12. doi:10.1109/IREP.2013.6629348.

[143] J.I. Sarasúa, G. Martínez-Lucas, M. Lafoz, Analysis of alternative frequency control schemes for increasing renewable energy penetration in El Hierro Island power system, Int. J. Electr. Power Energy Syst. 113 (2019) 807–823. doi:10.1016/j.ijepes.2019.06.008.

[144] G. Martínez-Lucas, J.I. Sarasúa, J.I. Pérez-Díaz, S. Martínez, D. Ochoa, Analysis of the implementation of the primary and/or inertial frequency control in variable speed wind turbines in an isolated power system with high renewable penetration. Case study: El hierro power system, Electron. 9 (2020) 1–24. doi:10.3390/electronics9060901.

[145] J.I. Sarasúa, G. Martínez-Lucas, J.I. Pérez-Díaz, D. Fernández-Muñoz, Alternative operating modes to reduce the load shedding in the power system of El Hierro Island, Int. J. Electr. Power Energy Syst. 128 (2021). doi:10.1016/j.ijepes.2020.106755.

[146] Z. Feng, F. Wei, C. Wu, Q. Sui, X. Lin, Z. Li, Novel Source-Storage Coordination Strategy Adaptive to Impulsive Generation Characteristic Suitable for Isolated Island Microgrid Scheduling, IEEE Trans. Smart Grid. 14 (2023) 3791–3803. doi:10.1109/TSG.2023.3244852.

[147] W. Zhou, C. Lou, Z. Li, L. Lu, H. Yang, Current status of research on optimum sizing of stand-alone hybrid solar-wind power generation systems, Appl. Energy. 87 (2010) 380–389. doi:10.1016/j.apenergy.2009.08.012.

[148] A. V. Ntomaris, S.I. Vagropoulos, A.G. Bakirtzis, Integration of a hybrid power station in the insular power system of Crete, in: IEEE PES Innov. Smart Grid Technol. Eur., IEEE, 2014: pp. 1–6. doi:10.1109/ISGTEurope.2014.7028875.

[149] D. Fernández-Muñoz, J.I. Pérez-Díaz, Unit commitment in a hybrid diesel/wind/pumped-storage isolated power system considering the net demand intra-hourly variability, IET Renew. Power Gener. 15 (2021) 30–42. doi:10.1049/rpg2.12003.

[150] D. Fernández-Muñoz, J.I. Pérez-Díaz, M. Chazarra, A two-stage stochastic optimisation model for the water value calculation in a hybrid diesel/ wind/pumped-storage power system, IET Renew. Power Gener. 13 (2019) 2156–2165. doi:10.1049/iet-rpg.2018.6151.

[151] R. López-Rodríguez, A. Aguilera-González, I. Vechiu, S. Bacha, Day-ahead mpc energy management system for an island wind/storage hybrid power plant, Energies. 14 (2021) 1–33. doi:10.3390/en14041066.

[152] G.E. Arnaoutakis, G. Kefala, E. Dakanali, D. Al Katsaprakakis, Combined Operation of Wind-Pumped Hydro Storage Plant with a Concentrating Solar Power Plant for Insular Systems: A Case Study for the Island of Rhodes, Energies. 15 (2022). doi:10.3390/en15186822.

[153] T. Nikolaou, G.S. Stavrakakis, K. Tsamoudalis, Modeling and optimal dimensioning of a pumped hydro energy storage system for the exploitation of the rejected wind energy in the non-interconnected electrical power system of the Crete Island, Greece, Energies. 13 (2020). doi:10.3390/en13112705.

[154] J.K. Kaldellis, M. Kapsali, K.A. Kavadias, Energy balance analysis of wind-based pumped hydro storage systems in remote island electrical networks, Appl. Energy. 87 (2010) 2427–2437. doi:10.1016/j.apenergy.2010.02.016.

[155] M. Kapsali, J.K. Kaldellis, Combining hydro and variable wind power generation by means of pumped-storage under economically viable terms, Appl. Energy. 87 (2010) 3475–3485. doi:10.1016/j.apenergy.2010.05.026.

[156] G.N. Perakis, D.A.I. Katsaprakakis, E.S. Karapidakis, Sizing a wind pumped storage hybrid power station for energy sufficiency of Leros' island, in: Synerg. MED 2022 - 2nd Int. Conf. Energy Transit. Mediterr. Area, Proc., Institute of Electrical and Electronics Engineers Inc., 2022. doi:10.1109/SyNERGYMED55767.2022.9941412.

[157] A.A. Agapitidou, S. Skroufouta, E. Baltas, Methodology for the Development of Hybrid Renewable Energy Systems (HRES) with Pumped Storage and Hydrogen Production on Lemnos Island, Earth (Switzerland). 3 (2022) 537–556. doi:10.3390/earth3020032.

[158] P.A. Dratsas, G.N. Psarros, S.A. Papathanassiou, D. Evagorou, A. Frixou, A. Poullikkas, Mid-term electricity storage needs of the power system of Cyprus, in: CIGRE Paris Sess., Paris, France, 2022: pp. 1–10.

[159] B. Roberts, Capturing grid power, IEEE Power Energy Mag. 7 (2009) 32–41. doi:10.1109/MPE.2009.932876.

[160] G.N. Psarros, A.G. Papakonstantinou, S.A. Papathanassiou, INTEGRATION OF STORAGE INTO LARGE ISLAND POWER SYSTEMS: THE CASE OF CYPRUS, in: 12th Mediterr. Conf. Power Gener. Transm. Distrib. Energy Convers. (MEDPOWER 2020), Institution of Engineering and Technology, 2021: pp. 270–275. doi:10.1049/icp.2021.1219.

[161] G. Edwards, S. Sheehy, C.J. Dent, M.C.M. Troffaes, Assessing the contribution of nightly rechargeable grid-scale storage to generation capacity adequacy, Sustain. Energy, Grids Networks. 12 (2017) 69–81. doi:10.1016/j.segan.2017.09.005.

[162] M.P. Evans, S.H. Tindemans, D. Angeli, Minimizing unserved energy using heterogeneous storage units, IEEE Trans. Power Syst. 34 (2019) 3647–3656. doi:10.1109/TPWRS.2019.2910388.

[163] R. Sioshansi, S.H. Madaeni, P. Denholm, A Dynamic Programming Approach to Estimate the Capacity Value of Energy Storage, IEEE Trans. Power Syst. 29 (2014) 395–403. doi:10.1109/TPWRS.2013.2279839.

[164] R. Billinton, W. Li, Reliability Assessment of Electric Power Systems Using Monte Carlo Methods, Springer US, Boston, MA, 1994. doi:10.1007/978-1-4899-1346-3.

[165] K. Mahmud, M.S. Rahman, J. Ravishankar, M.J. Hossain, J.M. Guerrero, Real-Time Load and Ancillary Support for a Remote Island Power System Using Electric Boats, IEEE Trans. Ind. Informatics. 16 (2020) 1516–1528. doi:10.1109/TII.2019.2926511.

[166] J. Gouveia, C.L. Moreira, J.A. Peças Lopes, Influence of Load Dynamics on Converter-Dominated Isolated Power Systems, Appl. Sci. 11 (2021) 2341. doi:10.3390/app11052341.

[167] T.T. Ku, C.S. Li, Implementation of Battery Energy Storage System for an Island Microgrid with High PV Penetration, IEEE Trans. Ind. Appl. 57 (2021) 3416–3424. doi:10.1109/TIA.2021.3075655.

[168] Y. Perez, F.J. Ramos Real, How to make a European integrated market in small and isolated electricity systems? The case of the Canary Islands, Energy Policy. 36 (2008) 4159–4167. doi:10.1016/j.enpol.2008.05.019.

[169] G.N. Psarros, Optimal operation of autonomous power systems with storage facilities for high RES penetration, (2019). doi:10.12681/eadd/46717.

[170] V. Rious, Y. Perez, Review of supporting scheme for island powersystem storage, Renew. Sustain. Energy Rev. 29 (2014) 754–765. doi:10.1016/j.rser.2013.08.015.

[171] P. Iliadis, S. Ntomalis, K. Atsonios, A. Nesiadis, N. Nikolopoulos, P. Grammelis, Energy management and techno-economic assessment of a predictive battery storage system applying a load levelling operational strategy in island systems, Int. J. Energy Res. 45 (2021) 2709–2727. doi:10.1002/er.5963.

[172] N. Hary, V. Rious, M. Saguan, The electricity generation adequacy problem: Assessing dynamic effects of capacity remuneration mechanisms, Energy Policy. 91 (2016) 113–127. doi:10.1016/j.enpol.2015.12.037.

[173] D.V. Pombo, J. Martinez-Rico, H.M. Marczinkowski, Towards 100% renewable islands in 2040 via generation expansion planning: The case of São Vicente, Cape Verde, Appl. Energy. 315 (2022). doi:10.1016/j.apenergy.2022.118869.

[174] D.V. Pombo, J. Martinez-Rico, S. V Spataru, H.W. Bindner, P.E. Sørensen, Decarbonizing energy islands with flexibility-enabling planning: The case of Santiago, Cape Verde, Renew. Sustain. Energy Rev. (2023). doi:10.11583/DTU.17.

[175] J. Hamilton, A. Tavakoli, M. Negnevitsky, X. Wang, Investigation of no load diesel technology in isolated power systems, in: 2016 IEEE Power Energy Soc. Gen. Meet., IEEE, 2016: pp. 1–5. doi:10.1109/PESGM.2016.7741324.

[176] J. Hamilton, M. Negnevitsky, X. Wang, S. Lyden, High penetration renewable generation within Australian isolated and remote power systems, Energy. 168 (2019) 684–692. doi:10.1016/j.energy.2018.11.118.

[177] A. Fernández-Guillamón, J.I. Sarasúa, M. Chazarra, A. Vigueras-Rodríguez, D. Fernández-Muñoz, Á. Molina-García, Frequency control analysis based on unit commitment schemes with high wind power integration: A Spanish isolated power system case study, Int. J. Electr. Power Energy Syst. 121 (2020)

106044. doi:10.1016/j.ijepes.2020.106044.

[178] E. Semshchikov, M. Negnevitsky, J. Hamilton, X. Wang, Cost-Efficient Strategy for High Renewable Energy Penetration in Isolated Power Systems, IEEE Trans. Power Syst. 35 (2020) 3719–3728. doi:10.1109/TPWRS.2020.2975236.

[179] A. Morales, D. Jara, J. Rengifo, D. Ochoa, Flywheel Energy Storage System to fast-frequency response provision on insular power systems from a microgrid perspective, in: 2022 IEEE Int. Conf. Ind. Technol., IEEE, 2022: pp. 1–6. doi:10.1109/ICIT48603.2022.10002748.

[180] D. Curto, S. Favuzza, V. Franzitta, A. Guercio, M. Amparo Navarro Navia, E. Telaretti, G. Zizzo, Grid Stability Improvement Using Synthetic Inertia by Battery Energy Storage Systems in Small Islands, Energy. 254 (2022) 124456. doi:10.1016/j.energy.2022.124456.

[181] T. Senjyu, D. Hayashi, A. Yona, N. Urasaki, T. Funabashi, Optimal configuration of power generating systems in isolated island with renewable energy, Renew. Energy. 32 (2007) 1917–1933. doi:10.1016/j.renene.2006.09.003.

[182] P. Meibom, R. Barth, B. Hasche, H. Brand, C. Weber, M. O'Malley, Stochastic Optimization Model to Study the Operational Impacts of High Wind Penetrations in Ireland, IEEE Trans. Power Syst. 26 (2011) 1367–1379. doi:10.1109/TPWRS.2010.2070848.

[183] P. Andrianesis, G. Liberopoulos, C. Varnavas, The impact of wind generation on isolated power systems: The case of Cyprus, in: 2013 IEEE Grenoble Conf., IEEE, 2013: pp. 1–6. doi:10.1109/PTC.2013.6652358.

[184] Y. Tian, L. Fan, Y. Tang, K. Wang, G. Li, H. Wang, A Coordinated Multi-Time Scale Robust Scheduling Framework for Isolated Power System with ESU under High RES Penetration, IEEE Access. 6 (2018) 9774–9784. doi:10.1109/ACCESS.2018.2792456.

[185] P. Nikolaidis, S. Chatzis, A. Poullikkas, Optimal planning of electricity storage to minimize operating reserve requirements in an isolated island grid, Energy Syst. 11 (2020) 1157–1174. doi:10.1007/s12667-019-00355-x.

[186] M. Hajjaji, D. Mezghani, C. Cristofari, A. Mami, Technical, Economic, and Intelligent Optimization for the Optimal Sizing of a Hybrid Renewable Energy System with a Multi Storage System on Remote Island in Tunisia, Electron. 11 (2022). doi:10.3390/electronics11203261.

[187] L. Heistrene, B. Azzopardi, A.V. Sant, P. Mishra, Stochastic Generation Scheduling of Insular Grids with High Penetration of Photovoltaic and Battery Energy Storage Systems: South Andaman Island Case Study, Energies. 15 (2022) 2612. doi:10.3390/en15072612.

[188] E. Rajaeian, R. Zeinali Davarani, Supplying the energy of islands with geopolitical positions using distributed generations to increase energy security, Int. J. Hydrogen Energy. (2023). doi:10.1016/j.ijhydene.2023.01.114.

[189] C. Berna-Escriche, C. Vargas-Salgado, D. Alfonso-Solar, A. Escrivá-Castells, Can a fully renewable system with storage cost-effectively cover the total demand of a big scale standalone grid? Analysis of three scenarios applied to the Grand Canary Island, Spain by 2040, J. Energy Storage. 52 (2022). doi:10.1016/j.est.2022.104774.

[190] D. Bionaz, P. Marocco, D. Ferrero, K. Sundseth, M. Santarelli, Life cycle environmental analysis of a hydrogen-based energy storage system for remote applications, Energy Reports. 8 (2022) 5080–5092. doi:10.1016/j.egyr.2022.03.181.

[191] R. Babaei, D.S.K. Ting, R. Carriveau, Optimization of hydrogen-producing sustainable island microgrids, Int. J. Hydrogen Energy. 47 (2022) 14375–14392. doi:10.1016/j.ijhydene.2022.02.187.

[192] A. Setas Lopes, R. Castro, C.S. Silva, Design of water pumped storage systems: A sensitivity and scenario analysis for island microgrids, Sustain. Energy Technol. Assessments. 42 (2020) 100847. doi:10.1016/j.seta.2020.100847.

[193] P. Benalcazar, A. Suski, J.K. Nski, Optimal Sizing and Scheduling of Hybrid Energy Systems: The Cases of Morona Santiago and the Galapagos Islands, Energies. 13 (2020). doi:10.3390/en13153933.

[194] J. Twitchell, K. DeSomber, D. Bhatnagar, Defining long duration energy storage, J. Energy Storage. 60 (2023) 105787. doi:10.1016/j.est.2022.105787.

[195] O.J. Guerra, Beyond short-duration energy storage, Nat. Energy. 6 (2021) 460–461. doi:10.1038/s41560-021-00837-2.